\documentclass{aa}  
\usepackage{listings}
\usepackage{graphicx}
\usepackage{txfonts}
\usepackage[multiple]{footmisc}

\begin{document}

   \title{The Intergalactic medium transmission towards z$\gtrsim$4 galaxies with VANDELS and the impact of dust attenuation\thanks{Based on observations made with ESO Telescopes at the La Silla or Paranal Observatories under programme ID(s) 194.A-2003}}
  \titlerunning{IGM towards z$\gtrsim$4 galaxies with VANDELS}
   \author{R. Thomas\inst{1}
           \and L. Pentericci\inst{\ref{RomeObs}} 
           \and O. Le Fevre\inst{\ref{LAM}}
           \and G. Zamorani\inst{\ref{BolognaObs}}
           \and D. Schaerer\inst{\ref{GeneveObs}}
           \and R. Amorin\inst{\ref{Serena1},\ref{Serena2}}
           \and M. Castellano\inst{\ref{RomeObs}}   
           \and A. C. Carnall\inst{\ref{SUPA}}
           \and S. Cristiani\inst{\ref{Triest}}
           \and F. Cullen\inst{\ref{SUPA}}
           \and S. L. Finkelstein\inst{\ref{Texas}}
           \and F. Fontanot\inst{\ref{Triest}}
           \and L. Guaita\inst{\ref{UC}}
           \and P. Hibon\inst{\ref{ESOChile}}
           \and N. Hathi\inst{\ref{STSI}}
           \and J. P. U. Fynbo\inst{\ref{Cosmicdawn}}
           \and Y. Khusanova\inst{\ref{LAM}}
           \and A. M. Koekemoer\inst{\ref{STSI}}           
           \and D. McLeod\inst{\ref{SUPA}}
           \and R. J. McLure\inst{\ref{SUPA}}
           \and F. Marchi\inst{\ref{RomeObs}}
           \and L. Pozzetti\inst{\ref{BolognaObs}}
           \and A. Saxena\inst{\ref{RomeObs}}
           \and M. Talia\inst{\ref{BolognaObs},\ref{BoloUni}}
           \and M. Bolzonella\inst{\ref{BolognaObs}}}

\institute{European Southern Observatory, Av. Alonso de C\'ordova 3107, Vitacura, Santiago, Chile \label{ESOChile}, \\
\email{rthomas@eso.org}
\and INAF, Osservatorio Astronomico di Roma, via Frascati 33, I-00078 Monteporzio Catone, Italy\label{RomeObs}
\and Aix Marseille Universit\'e, CNRS, LAM (Laboratoire d'Astrophysique de Marseille) UMR 7326, 13388, Marseille, France\label{LAM}
\and INAF - Osservatorio di Astrofisica e Scienza dello Spazio di Bologna, via Gobetti 93/3, I-40129, Bologna, Italy\label{BolognaObs}	
\and Observatoire de Gen\`eve, Universit\'e de Gen\`eve, 51 Ch. des Maillettes, 1290 Versoix, Switzerland\label{GeneveObs}
\and Instituto de Investigaci\'on Multidisciplinar en Ciencia y Tecnolog\'ia, Universidad de La Serena, Ra\'ul Bitr\'an 1305, La Serena, Chile\label{Serena1}
\and Departamento de F\'isica y Astronom\'ia, Universidad de La Serena, Norte, Av. Juan Cisternas 1200, La Serena, Chile\label{Serena2}
\and SUPA, Institute for Astronomy, University of Edinburgh, Royal Observatory, Edinburgh EH9 3HJ, UK\label{SUPA}
\and INAF-Astronomical Observatory of Trieste, via G.B.Tiepolo 11, 34143 Trieste, Italy\label{Triest}
\and Department of Astronomy, The University of Texas at Austin, Austin, TX, 78712\label{Texas}
\and Instituto de Astrof\'isica, Universidad Cat\'olica de Chile, Vicu\~na Mackenna 4860, Santiago, Chile \label{UC}
\and Space Telescope Science Institute, 3700 San Martin Drive, Baltimore, MD, 21218, USA\label{STSI}
\and The Cosmic Dawn Center, Niels Bohr Institute, Copenhagen University, Juliane Maries Vej 30, DK-2100 Copenhagen, Denmark \label{Cosmicdawn}
\and University of Bologna, Department of Physics and Astronomy (DIFA), Via Gobetti 93/2, I-40129, Bologna, Italy\label{BoloUni}
             }

   \date{;}

 
  \abstract
   {}
   {Our aim is to estimate the intergalactic medium transmission towards UV-selected star-forming galaxies at redshift 4 and above and study the effect of the dust attenuation on these measurements.}
   {The ultra-violet spectrum of high redshift galaxies is a combination of their intrinsic emission and  the effect of the Inter-Galactic medium (IGM) absorption along their line of sight. Using data coming from the unprecedented deep spectroscopy from the VANDELS ESO public survey carried out with the VIMOS instrument we compute both the dust extinction and the mean transmission of the IGM as well as its scatter from a set of 281 galaxies at z>3.87. Because of a degeneracy between the dust content of the galaxy and the IGM, we first estimate the stellar dust extinction parameter E(B-V) and study the result as a function of the dust prescription. Using these measurements as constraint for the spectral fit we estimate the IGM transmission Tr(Ly$\alpha$). Both photometric and spectroscopic SED fitting are done using the SPectroscopy And photometRy fiTting tool for Astronomical aNalysis (SPARTAN) that is able to fit the spectral continuum of the galaxies as well as photometric data.}
   {Using the classical Calzetti's attenuation law we find that E(B-V) goes from 0.11 at z=3.99 to 0.08 at z=5.15. These results are in very good agreement with previous measurements from the literature. We estimate the IGM transmission and find that the transmission is decreasing with increasing redshift from Tr(Ly$\alpha$)=0.53 at z=3.99 to 0.28 at z=5.15. We also find a large standard deviation around the average transmission that is more than 0.1 at every redshift. Our results are in very good agreement with both previous measurements from AGN studies and with theoretical models.}
   {}


   \maketitle
%

\section{Introduction}
\label{intro}

The observation of distant galaxies necessarily include the effect of the Inter-Galactic Medium (IGM) along the line of sight (LOS), and its associated extinction. The light coming from those sources is travelling through clouds that are lying along the line of sight. As the redshift of the source increases, the clouds along the LOS can be so numerous that all the light below the Lyman $\alpha$ line (at 1216\AA, hereafter, Ly$\alpha$) can be absorbed. Numerous authors have studied this phenomenon and it is thought that it is a natural result of the hierarchical formation of structure (e.g. \citealt{cen94}).

More than two decades ago, shortly after a work on the effect of the intergalactic medium on galaxy emission by \citet{Yoshii94},  \cite{madau95} (hereafter \textit{M95}) simulated the average IGM transmission as a function of redshift and found that it strongly decreases with increasing redshift. Moreover, the IGM leads to a very specific stair-like pattern where each step corresponds to a line of the Lyman series of the Hydrogen atom. In addition, a large scatter was expected and, for instance, the average transmission at z = 3.5 was estimated to range from 20\% to 70\% with an average of 40\% (M95). A decade later \cite{Meiksin06} (hereafter \textit{M06}) updated this model producing a new IGM prescription using the $\Lambda$-CDM model of \cite{Meiksin04}. It was found that the IGM transmission is higher than the one of M95, mainly because of differences in the estimates of the contributions of resonant absorption. More recently, \cite{Inoue14} developed a new model of transmission. Their model predicts a weaker absorption in the range z=3-5 than the M95 models while it becomes stronger at z>6.

For years, the average transmission (noted Tr(Ly$\alpha$)) has been estimated from the Ly$\alpha$ forest measurements on the LOS of QSOs. It is often referred to as the HI optical depth $\tau_{eff}$ with Tr(Ly$\alpha$) = $\exp(-\tau_{eff})$ and its measurements are used to constrain the intensity of the ionizing background (Haardt \& Madau 1996; Rauch et al. 1997; Bolton et al. 2005) and to investigate the sources responsible for the ionizing background. Surprisingly, only a few reports have been published on the observed dispersion in Tr(Ly$\alpha$) as a function of redshift.
Faucher-Giguère et al. (2008b) used 86 high-resolution quasar spectra with a high signal-to-noise ratio to provide reference measurements of the dispersion in Tr(Ly$\alpha$) over 2.2 < z < 4.6.

Until few years ago, no observational study had been made of the evolution of the IGM transmission from galaxy samples mainly because of the lack of large spectroscopic samples with high signal-to-noise ratios at high redshift that would probe a wavelength range significantly bluer than Ly$\alpha$. Hence, the comparison of IGM transmission towards extended galaxies with point-like QSOs had not yet been performed. 
In a recent paper (\citealt{Thomas17}) we were able to compute for the first time the IGM transmission towards a set of more than 2000 galaxies (with $\sim$120 of them at z>4) provided by the VIMOS Ultra Deep Survey (VUDS; \citealt{OLF15}). This study allowed us to show that, (i) the IGM transmission towards galaxies was a measurable parameter, (ii) the IGM transmission at $z<4$ was in very good agreement with the one computed towards QSO data in terms of both absolute measurements and also scatter around the mean values), (iii) at $z>4$ there might be a possible departure of the observational data from the theoretical prediction. This observed difference was interpreted as a signature of degeneracy between the dust and IGM models. 

In this paper we perform a study of 281 galaxies at z>4 from the very deep VANDELS survey \citep{VANDELS, Pent18} to compute the IGM properties. We therefore have more than twice the number of galaxies we had for the VUDS sample and with much deeper observation (ranging from 20 to 80 hours, instead of 14h). We also focus on the impact of different dust attenuation prescription on the IGM measurements. We describe the VANDELS galaxy sample and selection in Sect. \ref{Data} . The fitting method with the SPARTAN tool and the range of IGM templates used in the spectral fitting is described in Sect. \ref{SPARTAN} along the definition of the Ly$\alpha$ transmission we use in this paper. The estimation of the dust extinction and IGM transmission are described in Sect. \ref{sec_dust} and \ref{IGM_res}, respectively. We look at stacked spectra of different population in Sect.\ref{stacks}. Finally, we discuss the robustness of our results in section \ref{disc}.
All magnitudes are given in the AB system \citep{Oke83} and we use a cosmology with $\Omega_M$ = 0.3, $\Omega_{\Lambda}$ = 0.7 and h = 0.7.

\section{Data and sample selection}
\label{Data}
Our study is based on galaxies from the VANDELS survey. The data sample selection is described in \cite{VANDELS} while the data reduction and redshift measurements and validation are described in \cite{Pent18}. We briefly present an overview  of the survey in this section. 

VANDELS is a public spectroscopic survey carried out with the VIMOS instrument \citep{OLF03} located at the NASMYTH focus of the Unit Telescope 3 Melipal of the Very Large Telescope (VLT). It made use of the medium resolution grism spanning a wavelength window from 4800 to 10000\AA~with a spectral resolution of R=580. It targeted $\sim$2100 objects in a wide redshift range (1.0<z<7). 
Targets were selected in the two widely observed UDS and CDFS fields covering a total area of 0.2 deg$^{2}$. The primary target selection was performed using the photometric redshift technique. 
The reduction of the raw data was carried out using the EASYLIFE package \citep{EASYLIFE} and all redshifts were estimated using the EZ software \citep{Gari10}. A redshift flag has been assigned  to each redshift measurement. This flag corresponds to the probability of the redshift to be correct. The quality scheme is composed of six values. Flags 2, 3, 4 and 9 (for objects with a single emission line) are the most reliable flags with a probability to be correct of 75\%, 95\%, 100\% and 80\%, respectively. A quality flag of 1 indicates a 50\% probability of being correct, while a quality flag of 0 indicates that no redshift could be assigned. At the moment of writing, the internal VANDELS database provides 1527 unique sources (with more than 1300 available from the DR2). It gives access to 1-dimensional and 2-dimenstional spectra. 
Photometric data are available for each of the VANDELS galaxies from different ground-based or space-based observatory. Both fields are partially covered by optical and infrared photometric observations coming from the CANDELS survey with ACS and WFC3/IR and SPITZER/IRAC instruments \citep{Galametz13, Guo13}. Ground based data are also available with optical bands from the Subaru/Suprime-Cam instrument \citep{furusawa08, Cardamone10, furusawa16, Sobral12}, near infrared bands from the VIRCAM instrument from the VLT \citep{Jarvis13} and near infrared bands from the WIRcam camera of the CFHT \citep{Hsieh12}. We refer the reader to \cite{VANDELS} for further details.

\begin{figure*}[h!]
\centering
\includegraphics[width=9cm]{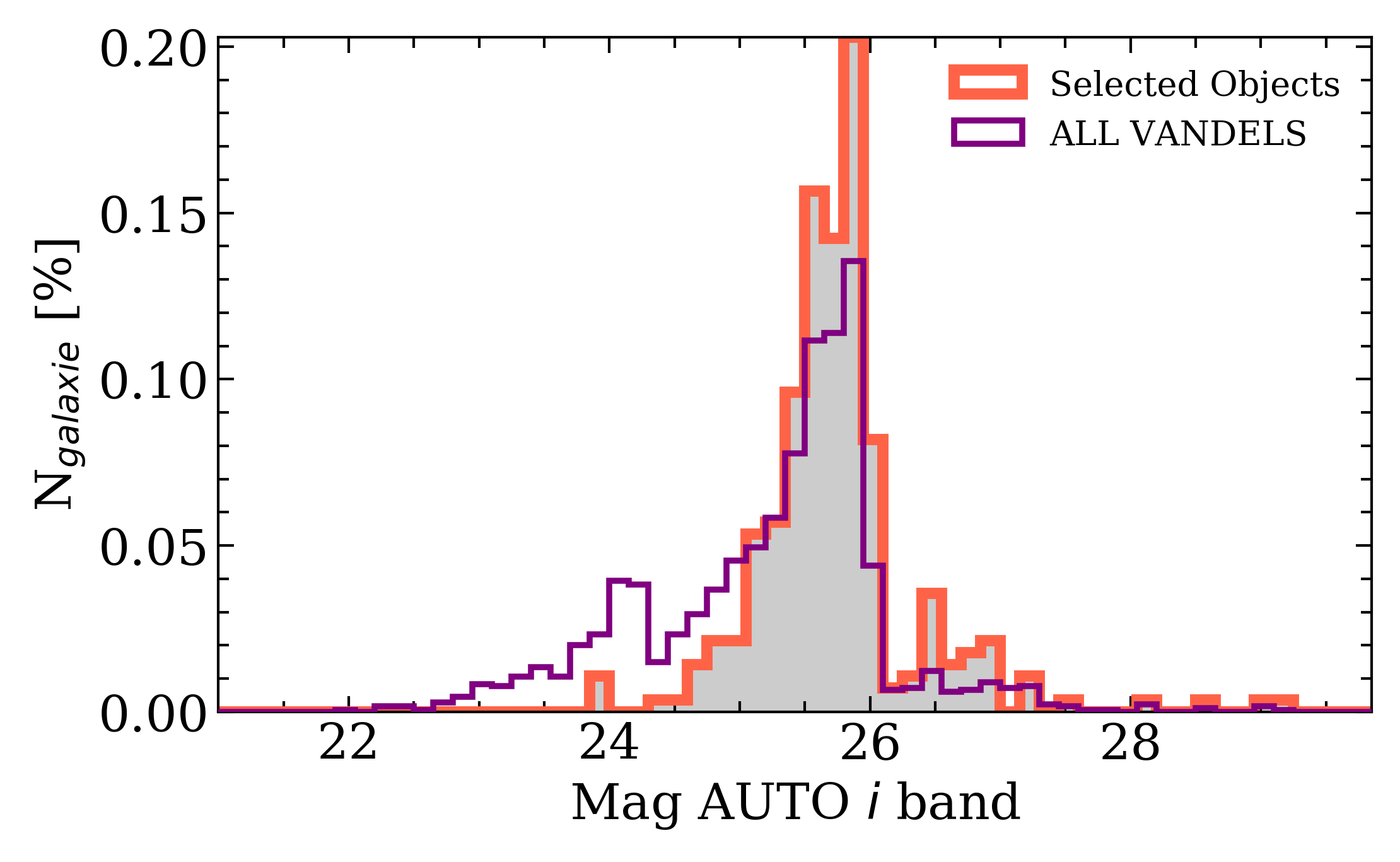}
\includegraphics[width=9cm]{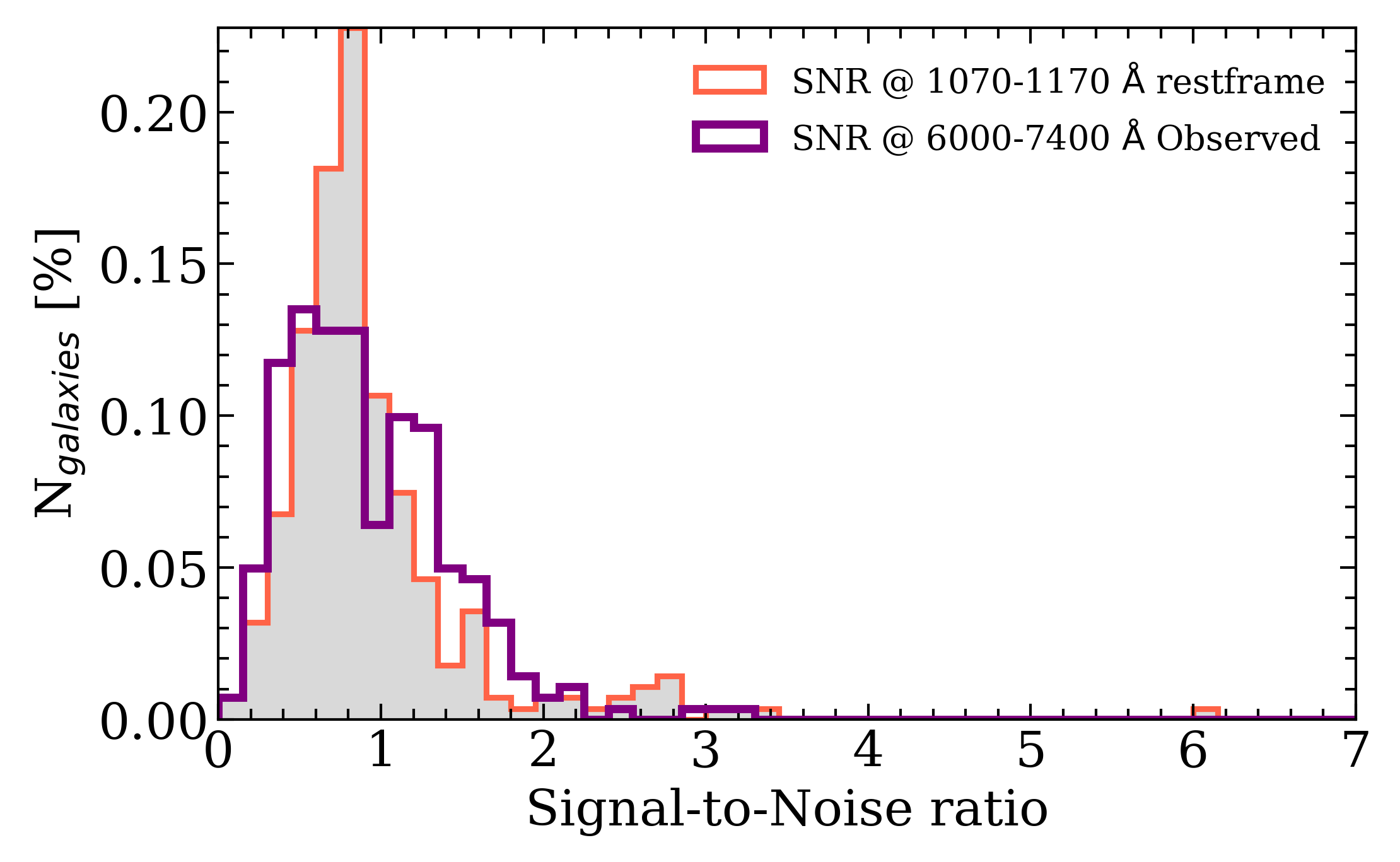}
    \caption{Observed properties of our selected sample of galaxies and comparison to the global released VANDELS data. \textit{Left:} Apparent magnitude (sextractor MAGAUTO) of our data in the \textit{i}-band. \textit{Right:} Signal-to-Noise measurements; at 1070-1170\AA~restframe in red and at a fixed observed wavelength (6000-7400\AA) in purple. In the former case the median SNR is 0.97.}
\label{SNR}
\end{figure*}

The aim of this paper is to study the IGM towards high redshift galaxies. As presented in Sect.\ref{intro}, the IGM signature in the spectra of distant galaxies is a stair-like pattern below the Ly$\alpha$ line. The Ly$\alpha$ transmission that we want to estimate is computed between the Ly$\alpha$ position, at 1216\AA, and the Ly$\beta$ position at 1025\AA. Therefore we must be able to observe this wavelength domain for our analysis. As the reduction process is sometimes not very efficient in extracting the edges of the spectra, we take a lower limit for our observed windows at 5000\AA~(instead of the nominal 4800\AA~limit of the medium resolution grism of VIMOS). This leads to a minimum redshift of z = 3.87. We do not impose, a-priori, any threshold on the signal to noise (SNR) nor the redshift flag for our working sample but we show the distribution ofSNR per spectral pixel measured with the recipe from \citealt{Stoehr08} (the dispersion of VANDELS spectra is $\sim$2.55\AA/pix) along the distribution of apparent magnitude in the \textit{i}-band in Fig.\ref{SNR}. This leads to a selected sample of 281 galaxies. In our sample 25 galaxies have redshift flag of 1, 69 have a redshift flag of 2 or 9, and 185 have a redshift flag of 3 or 4. Therefore 2/3 of our selected sample has an assigned redshift with a probability to be correct higher than 95\%. The stability of our results with respect to the choice of redshift flag is discussed in Sect.\ref{disc}.

\section{Method}
\label{SPARTAN}

\subsection{The SPARTAN tool}
To estimate the IGM transmission toward our galaxies we use the SPARTAN tool which is able to fit both photometry and spectroscopic data. In this paper we use the capability of SPARTAN to fit spectroscopic dataset and photometric dataset separately. This single data type fitting follows the same recipe as other codes used in the literature (e.g. \citealt{Salim07}, \citealt{Thomas17age}). For a given object and a single template the $\chi^{2}$ and associated probability are estimated with:

\begin{equation}
\chi^{2} = \sum_{i=1}^{N}\frac{(F_{obs,i}-A_{i}F_{syn,i})^{2}}{\sigma_{i}^{2}};\;\; P=\exp\left[-\frac{1}{2}(\chi^{2}-\chi^{2}_{min}) \right]
\label{eq1}
\end{equation}

where N, F$_{obs,i}$, F$_{syn,i}$, $\sigma_i$ ,A$_i$ and $\chi^{2}_{min}$ stand for the number of observed data points, the flux of the data point itself, the synthetic template value at the same wavelength, the observed error associated to F$_{obs,i}$, the normalization factor applied to the template, and the minimum $\chi^{2}$ of the library of template, respectively. The latter is used to set the maximum of the probability distribution to unity. From the properties of the exponential function this is only a normalization factor and does not change the values of the parameters' estimation nor their errors. The set of probability values (second part of equation \ref{eq1}) are then used to create the probability distribution function (PDF, whose integral is normalized to unity) for each parameter to be estimated. From the PDF we create the cumulative distribution function (CDF) where the measured value of the parameter is taken where CDF(X)=0.5 and the errors on this measurement correspond to the value of the parameter for which the CDF=0.05 and 0.95.\\

The photometric fitting process is performed as follows. The set of synthetic templates is redshifted to the redshift of the fitted galaxies and then normalized in one pre-defined band. For the photometric fitting we performed in this paper this normalization is applied in the \textit{i}-band. Once this normalization is done, SPARTAN convolves the normalized template with all the photometric bandpasses available for the observed galaxy. Finally, the relations in Eq.\ref{eq1} are applied to estimate the physical parameters of the observed galaxy and their associated errors. \\

When dealing with spectroscopic data, the general principle of the fitting process is similar. Nevertheless, this type of data allows for a different normalisation method.  SPARTAN has to normalize the redshifted template to the observed spectrum. As for the photometry, we can consider a photometric filter and estimate the magnitude in the same filter from the spectrum itself. This magnitude serves as a normalization to all the templates. This approach, widely used in the literature with photometric dataset, uses normalization that is always done in a given photometric band (e.g., the i-band). As a result, each galaxy is normalized to the template in a different rest-frame region and all the galaxies are not treated in a similar manner. The spectroscopy opens a new \textit{redshift-dependent} method of normalization. This method uses an emission line free region available in the spectrum. In the UV spectrum of distant galaxies, a region free from strong spectral line is between 1070 and 1170 \AA~(rest-frame). When fitting a UV spectrum at z=4.5, this spectral region will be shifted at 5885-6435\AA. SPARTAN will compute a  spectro-photometric point in this region directly in the template and in the data using a box filter of the size of this region. This \textit{box-magnitude} will then be used to normalize the template to the observed spectrum. At higher redshift, i.e. z=5.0, this spectral region will be at a redder wavelength (6420-7020\AA) and this observed-frame region will be used to perform the normalization as well. This new method of normalization has the merit of being consistent from one object to another. Moreover, as it is used in emission line free region, it relies less on the emission line physics of the templates. We use in this paper the latter redshift dependent normalisation method and we make the normalisation in the region 1070-1170\AA~(restframe). This choice is supported by; (i) Once redshifted this region provides a wide window for SNR estimation ($\sim$500\AA~@ z=3.87 and $\sim$750\AA~@ z=6.5), (ii) It is free of strong emission line, (iii) considering the VIMOS wavelength window, it is one of the only wide-enough wavelength range available across the redshift range we consider.

\subsection{IGM models and Tr(Ly$\alpha$) definition}
\begin{figure}[h!]
\centering
\includegraphics[width=\hsize]{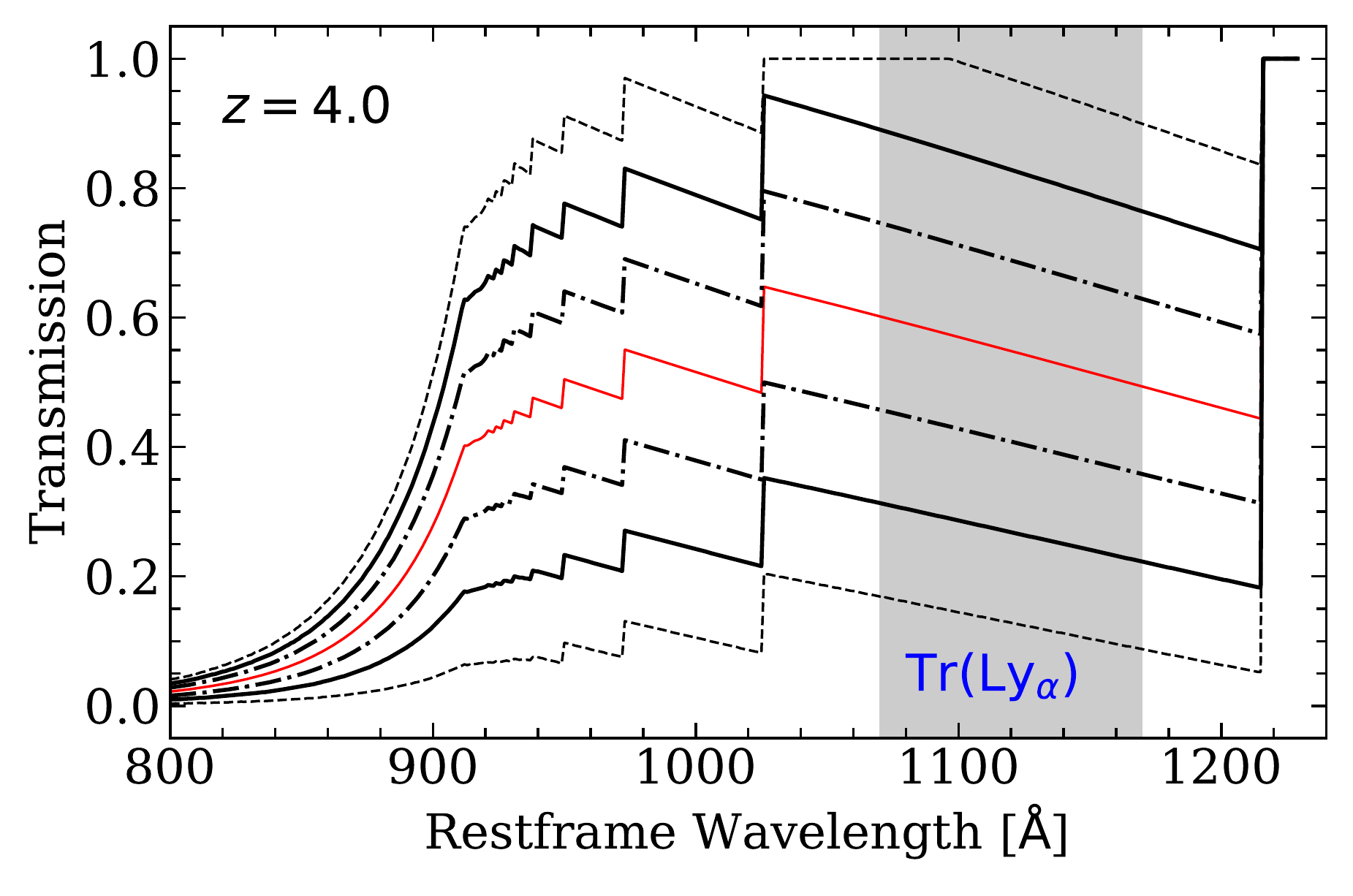}
\caption{Example of IGM transmission curves at z=4.0. The red curve is from M06 Prescription while the black curves represent the augmented prescription from \cite{Thomas17}. The latter allows, at this redshift to span possible transmission from $\sim$15\% to $\sim$90\%. The grey area shows where we compute the Ly$\alpha$ transmission.}
\label{trdef}
\end{figure}

To estimate the IGM transmission we must be able to fit it. For years, the IGM transmission was fixed at a single value at a given redshift most often using the M95 model that provides a single transmission curve at a given redshift. Therefore, it was assumed that at a given redshift, the lines of sight of objects observed at a different position in the sky are populated by hydrogen clouds with the same properties. In M95, the author provides an estimate of the 1 $\sigma$ dispersion and, as mentioned in Sect. \ref{intro}, it can vary from 20\% to 70\% at z=3.5. Additionally, it was shown that this dispersion around the mean IGM could produce better photometric redshift \citep{Furusawa00}. Therefore, we proposed in our previous paper to use a set of empirical models that can reproduce this dispersion in the IGM transmission (\citealt{Thomas17}, hereafter T17). We summarize here how these templates were constructed. To test different line of sights during the SED-fitting we constructed 6 additional templates around the mean of M06. This additional models were built considering the $\pm1$ sigma variation of M95 IGM models (see Fig.3a in M95 paper at $z=3.5$) that we propagated at any redshift. Finally, to explore more possibilities, we created, still from this $\pm1$ sigma variation, the $\pm0.5\sigma$ and $\pm1.5\sigma$. As a result, the IGM can be chosen from the set of 7 discrete values at any redshift and 
this allows us to use the IGM as a free parameter in out fitting procedure and explore a larger range of IGM transmission. At z=3.0 it ranges from 20\% to 100\% while at z=5.0, it ranges from 5\% to 50\%. As an example, we show in Fig.\ref{trdef} the set of extinction curves at z=4.0.
In this paper we aim at computing the Ly$\alpha$ Transmission, Tr(Ly$_\alpha$) which is determined as the mean transmission between 1070\AA~and 1170\AA~computed on the transmission curve itself, shown in Fig.\ref{trdef} by the grey area. In the case of SPARTAN we use the PDF of Tr(Ly$_\alpha$) to estimate the value of the parameter, as described in Sect. \ref{SPARTAN}.

Finally, we emphasize that the IGM models we use here, while based on simulation, are also empirical (in the additional curves we use). More recent models include more components in the simulations such as the inclusion of the CGM \citep{Steidel18,Kakiichi18}. We will compare these different prescriptions in a near-future paper. It is also worth noting that the general shape of the curves is the same from one model to another while it can vary from one line-of-sight to another, depending on the presence of absorbers. In T17 we compared the results of the fit using the templates presented here and real Lyman $\alpha$ forest simulation \citep{Bautista15} and found that there was a very good agreement in the resulting measurement of the Ly$\alpha$ Transmission.

\section{Dust content of z>4 galaxies}
\label{sec_dust}

\begin{figure}[h!]
\centering
\includegraphics[width=9cm]{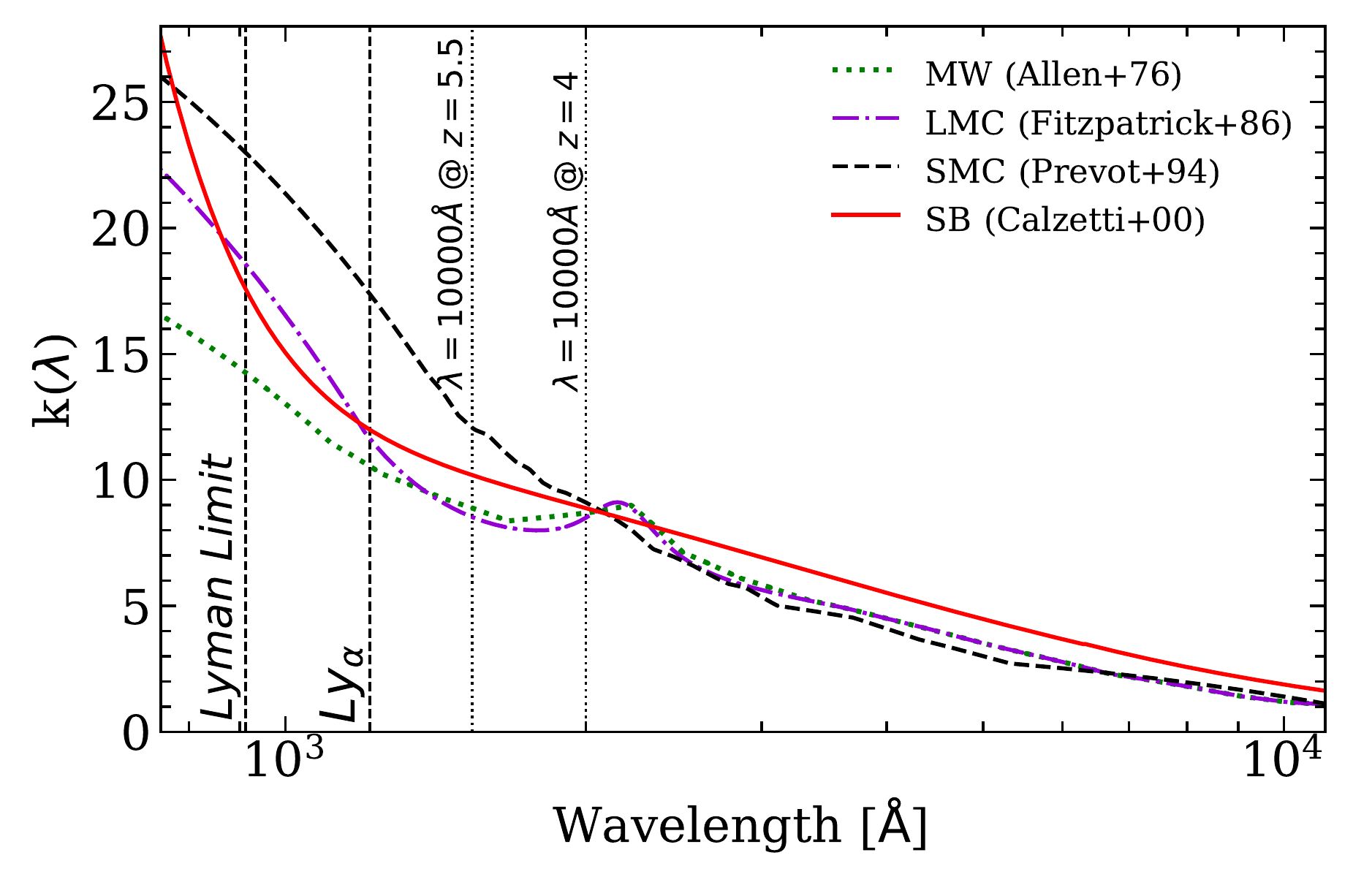}
\includegraphics[width=9cm]{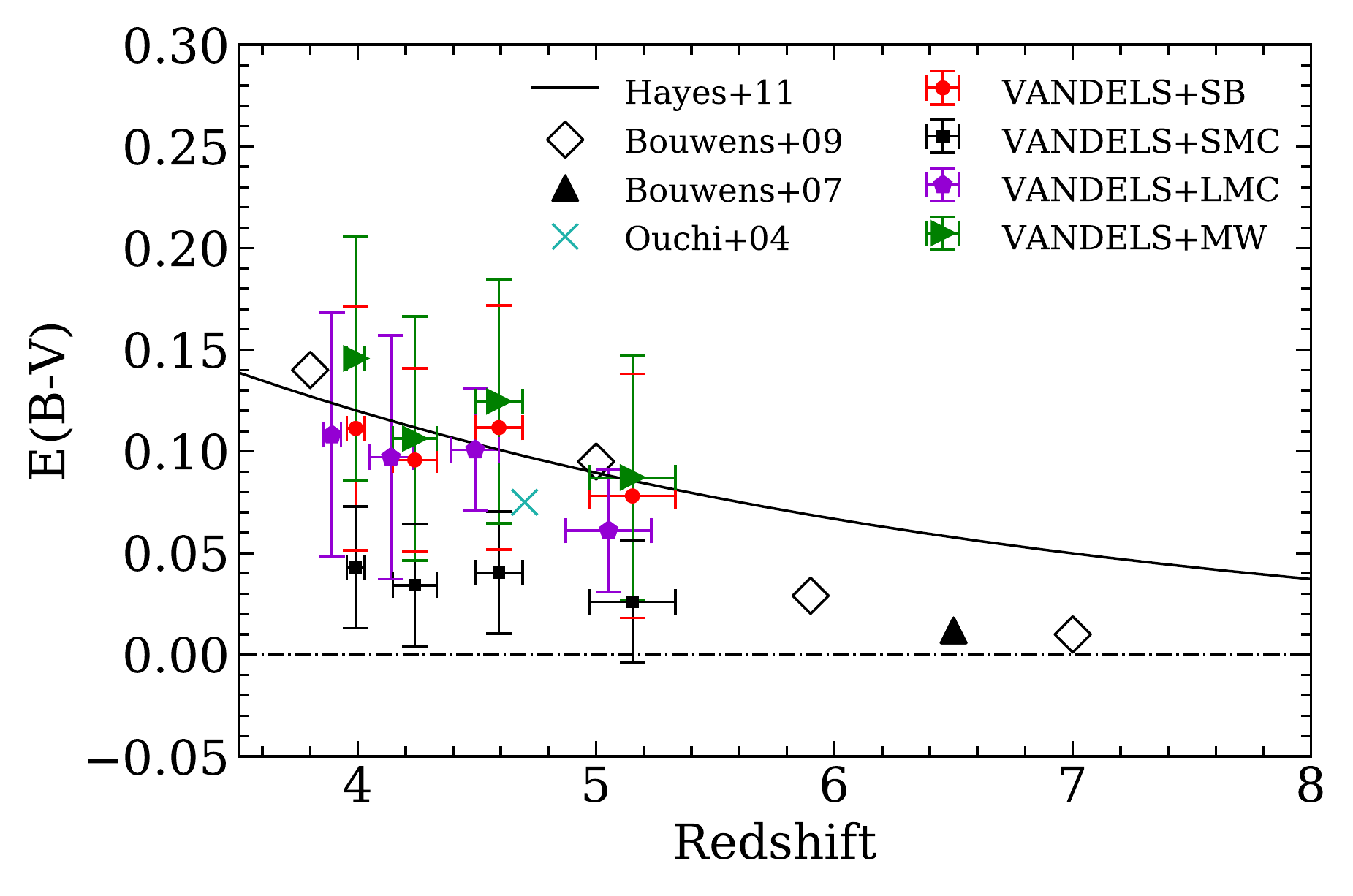}
\caption{Dust evolution from our 281 galaxies. \textit{Top Panel:} The four different dust prescriptions used to estimate the dust extinction in our sample. In red we show the prescription from \cite{Calzetti00}, in black the prescription from \cite{prevot} for the SMC, in violet the prescription from \cite{fitz} for the LMC and in green for the Milky Way prescription by \cite{Allen}. \textit{Bottom Panel:} Evolution with redshift of the dust attenuation in our selected sample of 281 galaxies from the photometric fitting for the four dust prescription shown in the Top Panel. Measurements report the mean and median absolute deviation for both the redshift and the E(B-V) values. We compare our results with previous measurements found in the literature at similar redshifts. The empty black diamonds are estimation from \cite{Bowens09}, black triangles from \cite{Bouwens07} and light blue cross from \cite{Ouchi04}. The dashed black line shows a fit from \cite{Hayes11}. \textit{Note:} All the VANDELS estimation are at the same redshifts, violet ones have been slightly shifted for clarity.}
\label{dust}
\end{figure}

In \cite{Thomas17} we identified a potential strong degeneracy between the estimates of the dust content of the galaxy and the IGM transmission prescription. This degeneracy is more prominent at z>4. In other words, the same data can be fitted with high values of both dust extinction and IGM transmission or lower values for both parameters. This is due to the small wavelength range available from UV spectra for fitting that is not able to constrain the dust content of the galaxy. In order to address this problem in the present work, we measure the IGM transmission in a two step process. First, we estimate the dust content of each galaxy in our sample using the photometric data presented in Sect. \ref{Data}. We fit the SED over a broader wavelength range than the spectra including NIR data, providing robust constraints on dust extinction. Then, we estimate Tr(Ly$\alpha$), keeping the E(B-V) value fixed to the one measured during the photometric-fitting process.

For this two-step fitting process we use the following parameter space. We use \cite{BC03} models with a \cite{Chab03} initial mass function. The stellar-phase metallicity ranges from sub-solar (0.2$Z_{\odot}$ and 0.4$Z_{\odot}$) to solar (1.0$Z_{\odot}$). We assume a star formation history prescription that is exponentially delayed with a timescale parameter, $\tau$, ranging from 0.1 Gyr to 1.0 Gyr. The ages range from 1 Myr to 3 Gyr in 24 steps. It is worth noting that this range of age is further limited by the age of the Universe at the redshift that is considered during the fit. The emission lines are added to the template following the work of \cite{Daniel09} that adds nebular continuum and emission using the conversion from ionizing photon into H$\beta$ luminosity. Then other emission lines are added using line ratio from \cite{AndersFritz03}.
This first step is made to estimate the dust extinction. Therefore, we use in this section four different dust extinction curves\footnote{ http://webast.ast.obs-mip.fr/hyperz/hyperz$\_$manual1/node10.html}: 
the classical starburst galaxy prescription from \cite{Calzetti00} (hereafter \textit{SB}) with an extrapolation, the Small Magellanic Cloud from (\citealt{prevot}, hereafter \textit{SMC}), the prescription for the Large Magellanic Cloud (\citealt{fitz}, hereafter \textit{LMC}) and finally the prescription for the Milky Way (\citealt{Allen}, hereafter \textit{MW}). All the curves are presented in Fig. \ref{dust} (top panel). For the photometric fitting, the E(B-V) parameter can vary from 0.0 to 0.39 (in 0.03 steps). Finally, the IGM prescription is using the models developed in \cite{Thomas17} based on the M06 models (see previous section). The redshift used for this fitting is the spectroscopic redshift, $z_{spec}$. Finally, it is worth noting that during this fitting process the IGM is estimated. These IGM measurements from the pure photometric fitting are discussed in Sect.\ref{disc}.

\begin{table}
\centering
\caption{Measurements of E(B-V) from the fit of photometry only using four different dust prescriptions. Each value represents the mean in each redshfit bins and the errors is the median absolute deviation.}
\begin{tabular}{ccccc}
\hline 
<z> & MW & LMC & SMC & SB  \\ 
\hline 
3.99 & 0.14$\pm$0.06 & 0.11$\pm$0.06 & 0.04$\pm$0.03 & 0.11$\pm$0.06  \\ 
4.23 & 0.11$\pm$0.06 & 0.09$\pm$0.06 & 0.03$\pm$0.03 & 0.10$\pm$0.05  \\ 
4.59 & 0.12$\pm$0.06 & 0.08$\pm$0.03 & 0.04$\pm$0.03 & 0.11$\pm$0.06  \\ 
5.15 & 0.08$\pm$0.06 & 0.05$\pm$0.03 & 0.03$\pm$0.03 & 0.08$\pm$0.06  \\ 
\hline 
\end{tabular} 
\label{tabledust}
\end{table}
The dust extinction measurements can be seen in Fig. \ref{dust} (bottom panel) that shows the evolution of the dust attenuation with redshift for our 281 selected galaxies and in Tab. \ref{tabledust} we report the measurements. Using the SB prescription we report mean values of E(B-V)=0.11; 0.10; 0.10 and 0.08 at z=3.99; 4.23; 4.59 and 5.15, respectively. Using the \textit{LMC} curve the measurements are very similar and we obtain 0.11, 0.09, 0.08 and 0.05 at the same redshifts. The prescription of the \textit{SMC} leads to a much weaker extinction at any redshift with 0.04, 0.03, 0.03 and 0.02 while the Milky-Way extinction from MW leads to slightly higher values of E(B-V with 0.14, 0.11, 0.12 and 0.08. This is easily explained by the fact that for a fixed value of E(B-V) the curve of the \textit{SMC} will lead to a much stronger extinction when studying UV-restframe galaxies (below 2000\AA) while the curve from the \textit{MW} will give lower extinctions. The measurements obtained using \textit{SB}, \textit{LMC} and \textit{MW} are in good agreement with previous estimation at similar redshifts. Studying dropout selected galaxies \cite{Bowens09} reports an E(B-V) measurement of 0.14 at z$\sim$3.8 while E(B-V)=0.095 at z=5.0. At similar redshift, \cite{Ouchi04} used Lyman break galaxies between 3.5<z<5.2 and measured E(B-V)=0.075 at z=4.7. It is worth mentioning that SB-like laws have been supported by other studies in the literature (e-g \citealt{McLure18}). On the contrary, the measurements using \textit{SMC} are in strong disagreement with previous measurement from the literature, as reported in \cite{scoville15} and \cite{Fudamoto17}. We will use, in the rest of the paper, both \textit{SB} and \textit{MW} models (\textit{LMC} being very close to the \textit{SB}) and see how this influences the measurements of the IGM.

\section{Tr(Ly$\alpha$) towards z>4 galaxies}
\label{IGM_res}

\begin{figure*}[h!]
\centering
\includegraphics[width=8cm]{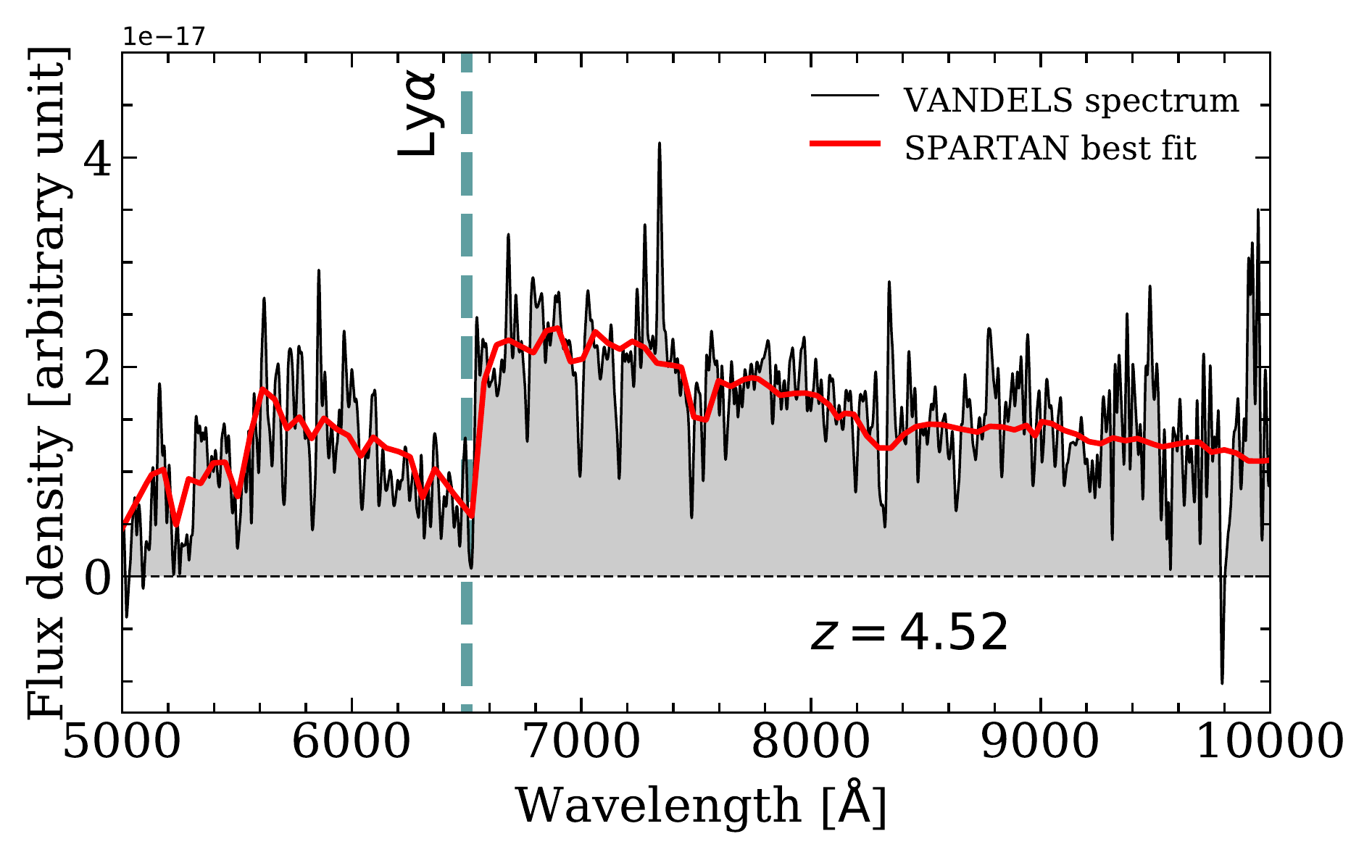}
\includegraphics[width=8cm]{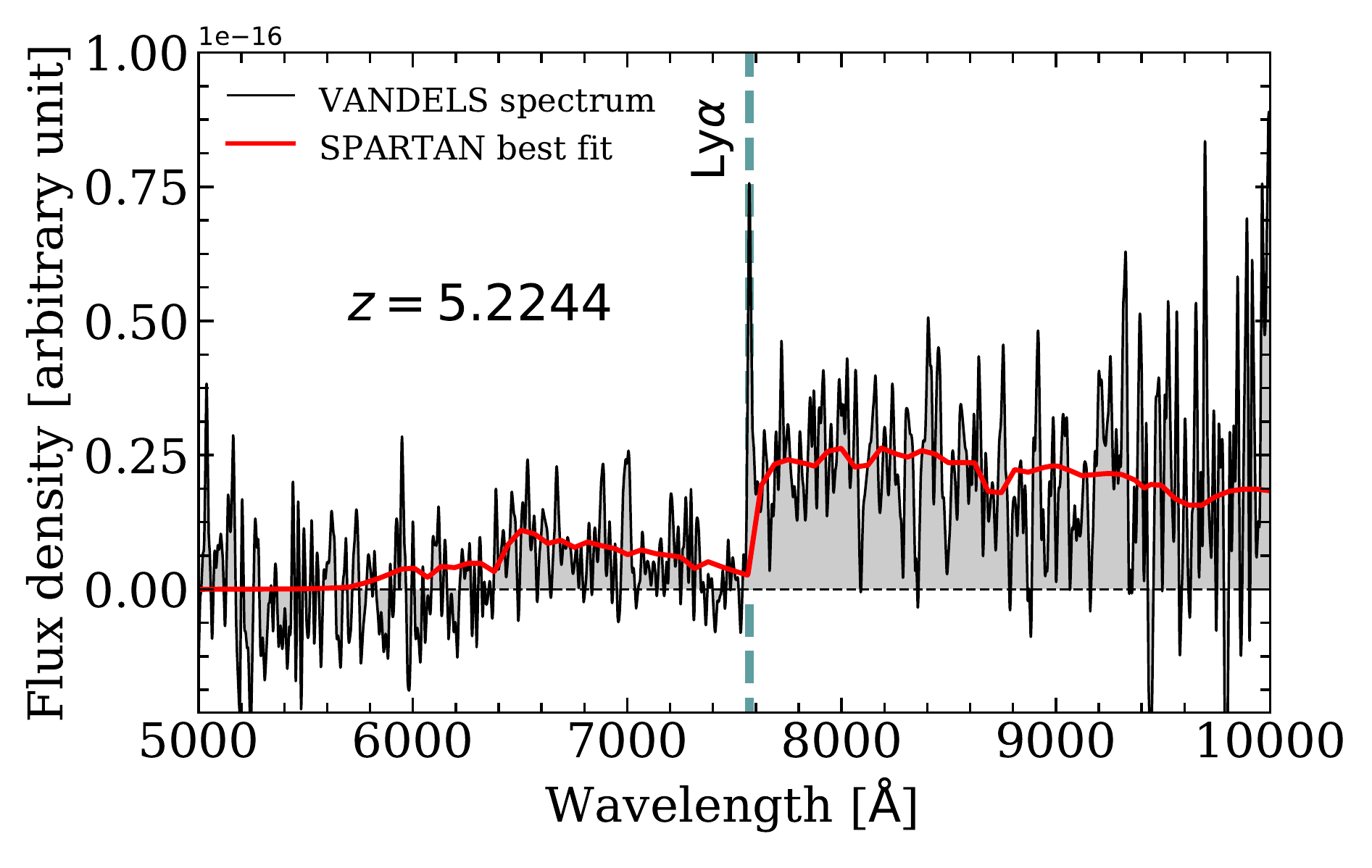}
\includegraphics[width=8cm]{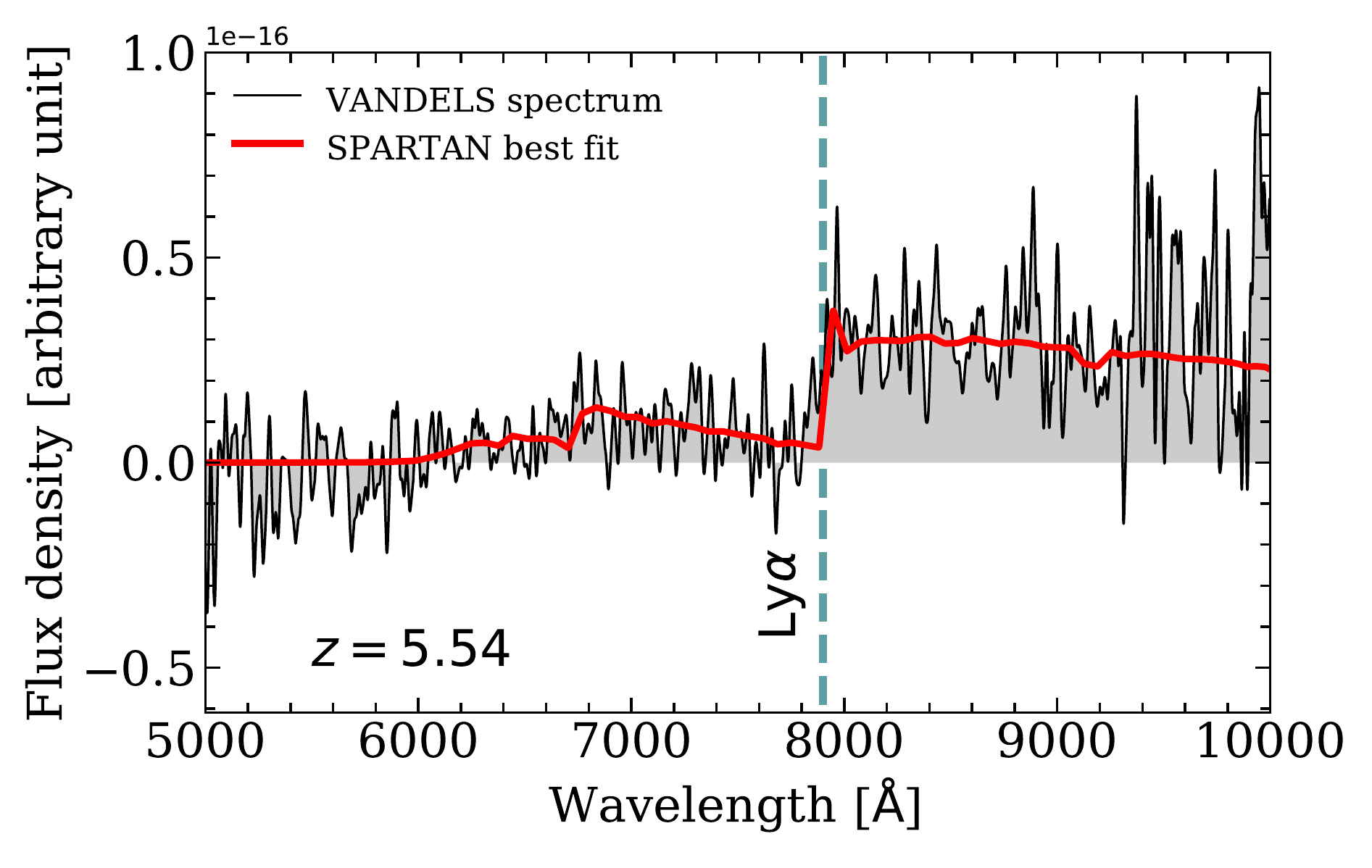}
\includegraphics[width=8cm]{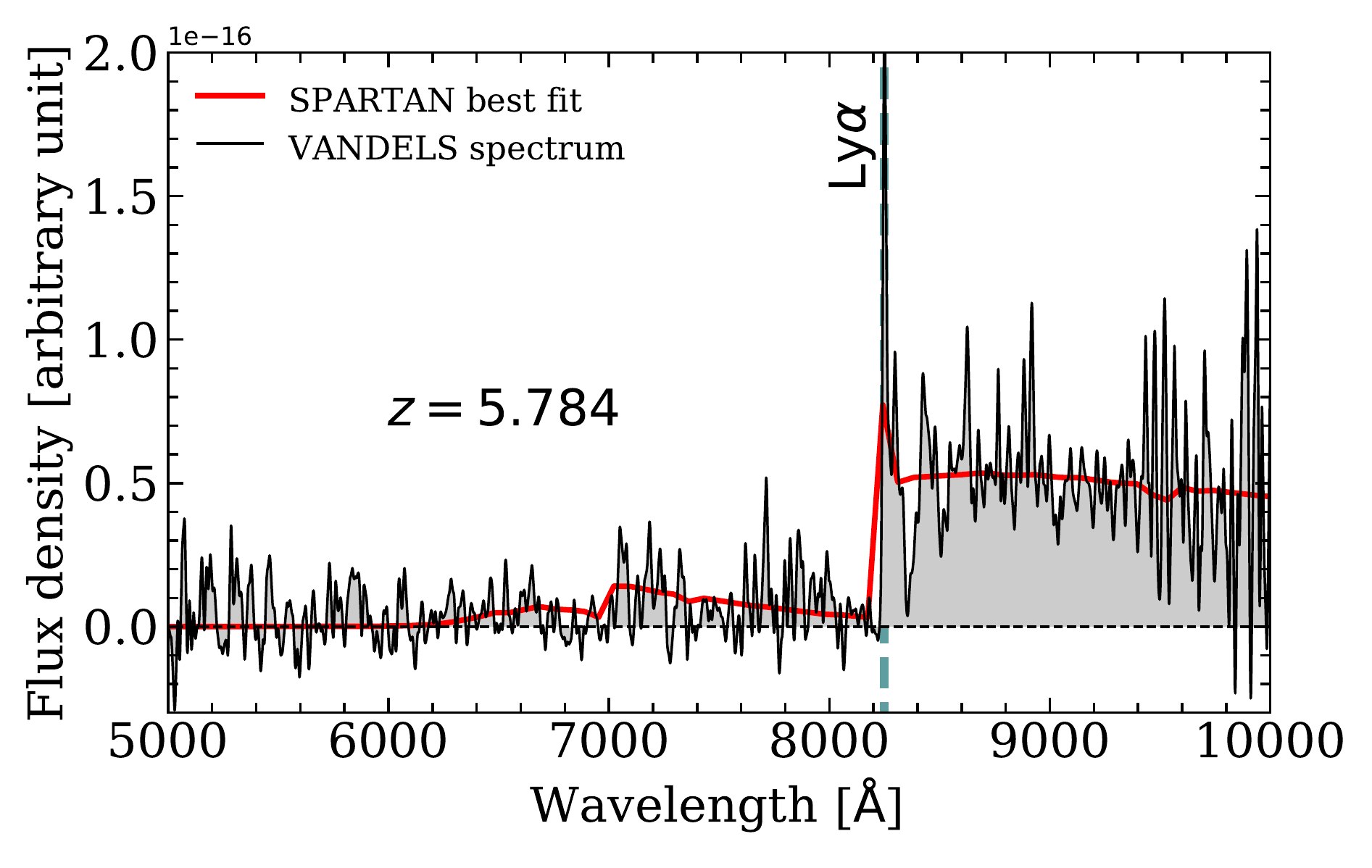}
\caption{Example of fits of VANDELS galaxies at redshift $z>4$. In each plot the spectrum is shown by the black line while the best fit, produced by SPARTAN, is given in red. We indicate the position of the Ly$\alpha$ line and the redshift for each spectrum.}
\label{specfit}
\end{figure*}
We measured in the last section the dust content of our galaxies. We now move towards the estimation of the IGM transmission with the spectral fit. In order to estimate it we constrain the spectral fit of our VANDELS spectroscopic data fixing the E(B-V) to the one measured during the fit of the photometric data. We consider individual E(B-V) value for each of our galaxies and do not use the average values presented in Fig. \ref{dust}. The other parameters, such as age or metallicity are still free to vary and the parameter ranges correspond to the ones of the fit of Sect.\ref{sec_dust}. Examples of spectral fit of VANDELS galaxies are presented in Fig. \ref{specfit} at various redshift and they show that SPARTAN reproduces very well the UV continuum of our galaxies at all wavelengths. It is worth mentioning that we can see that the Ly$\alpha$ is poorly reproduced. As presented in the previous section, the emission lines are added using line ratios and therefore not fitted individually. Our results remain if we mask out the line during the fit.
\begin{table*}
\centering
\caption{Measurements of the Tr(Ly$\alpha$) from our study. We report in this table the redshift, the Tr(Ly$\alpha$), the standard deviation, the standard error, the Tr(Ly$\alpha$) for redshift flag 2, 3 and 4 only (with the number of galaxies with these flags in parenthesis) and the measurements of the same quantity from the pure photometric fitting.}
\begin{tabular}{ccccccc}
\hline 
Redshift &N$_{gal}$ & Tr(Ly$\alpha$) & Std deviation & Standard error & Flag 2,3\&4 & Photometry \\ 
\hline 
3.99 & 81 & 0.55 & 0.14 & 0.015 & 0.54 (58) & 0.41 \\ 
4.23 & 74 & 0.49 & 0.14 & 0.016 & 0.48 (50)& 0.24 \\ 
4.59 & 72 & 0.42 & 0.13 & 0.015 & 0.41 (54)& 0.34 \\ 
5.15 & 54 & 0.29 & 0.11 & 0.015 & 0.30 (23)& 0.26 \\ 
\hline 
\end{tabular} 
\label{table}
\end{table*}

\begin{figure}[h!]
\centering
\includegraphics[width=4.45cm]{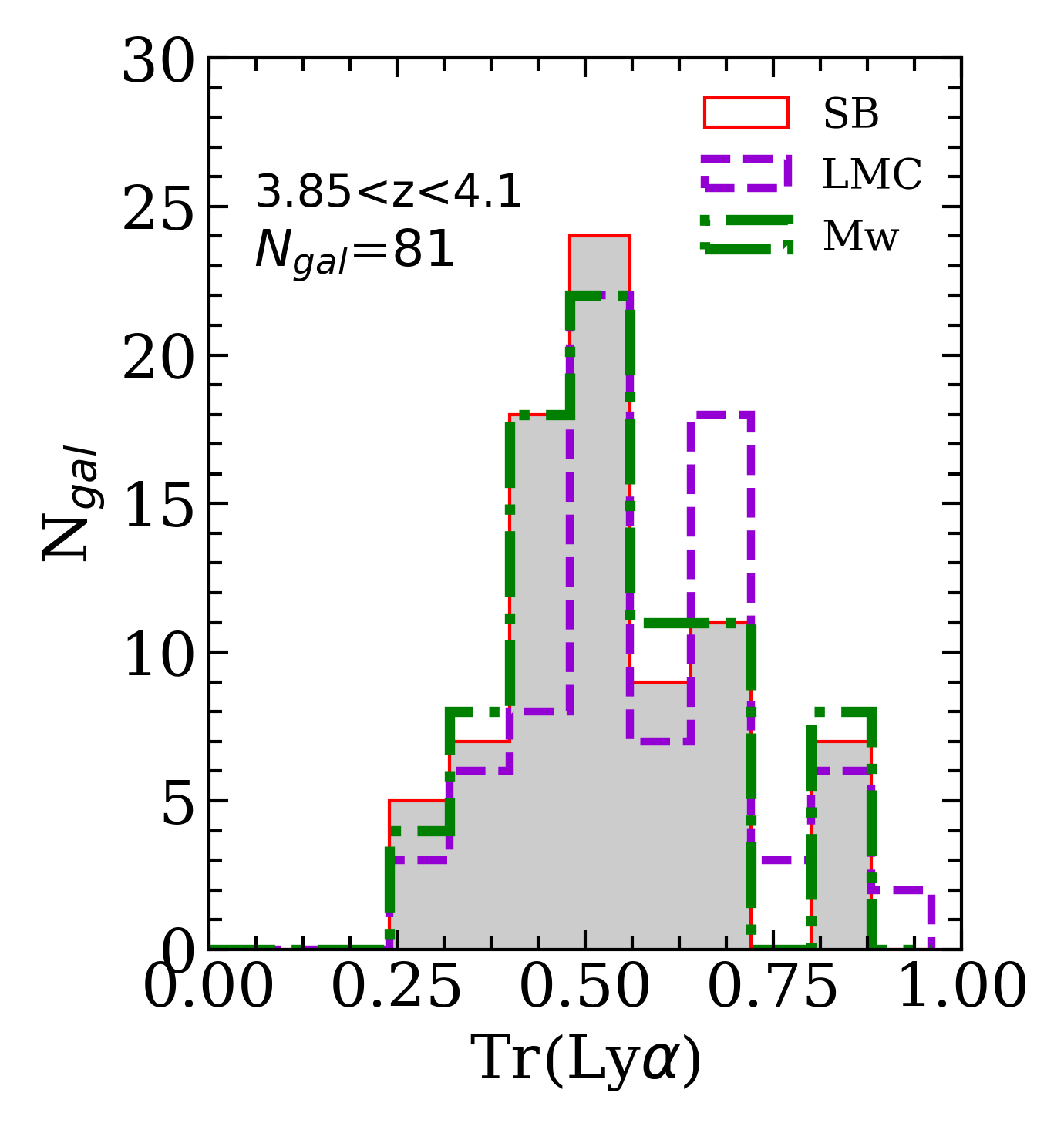}
\includegraphics[width=4.45cm]{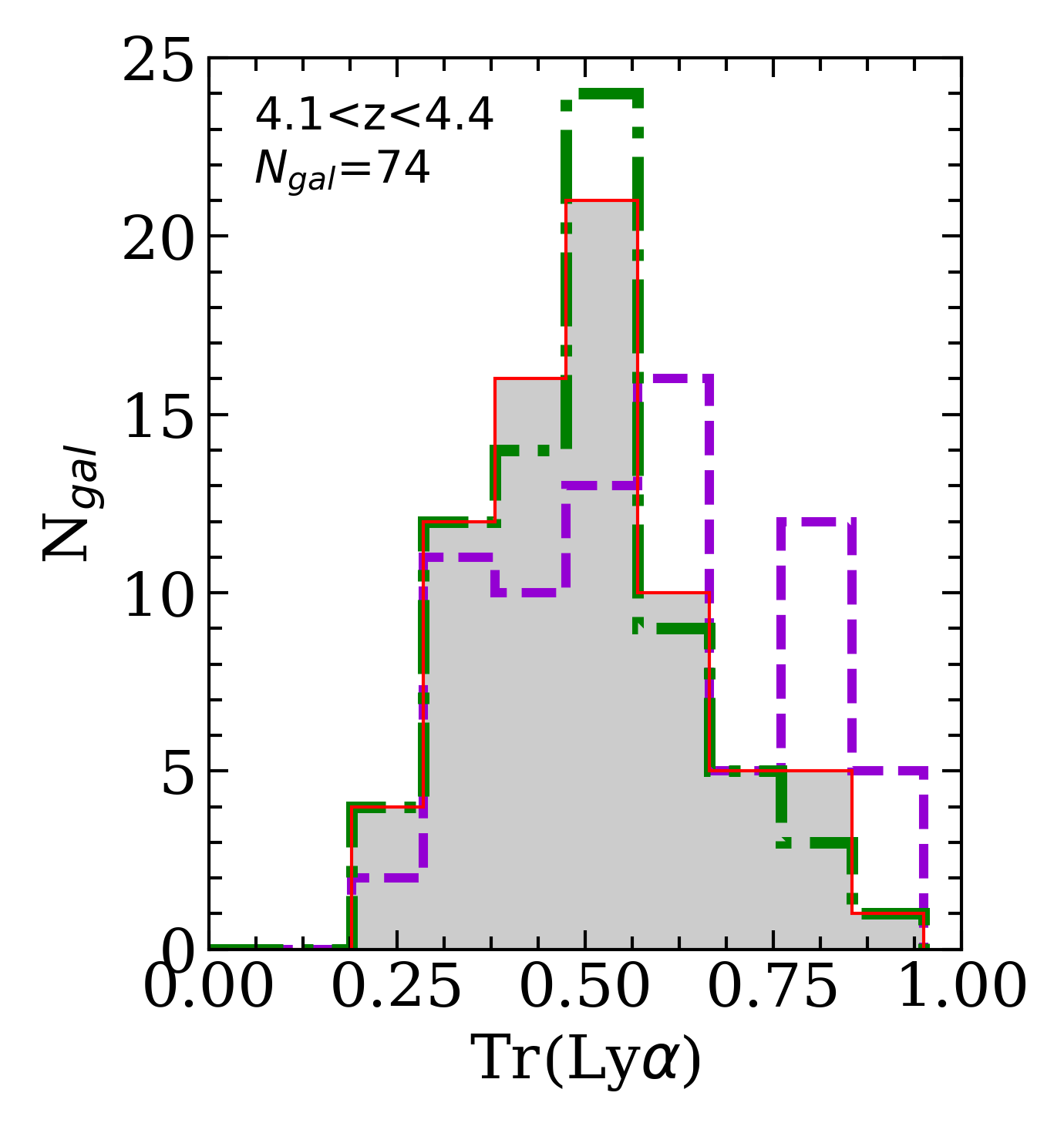}
\includegraphics[width=4.45cm]{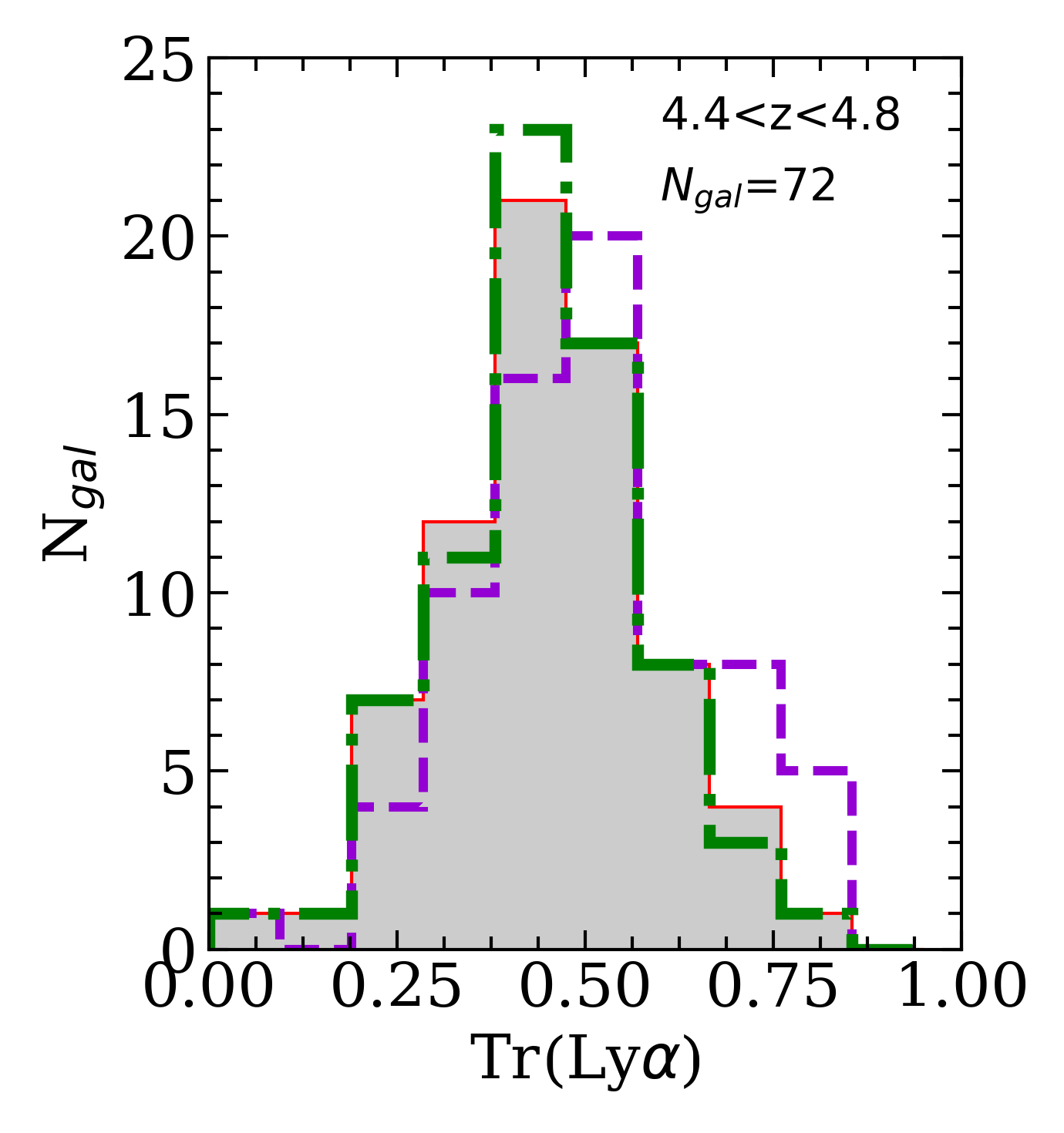}
\includegraphics[width=4.45cm]{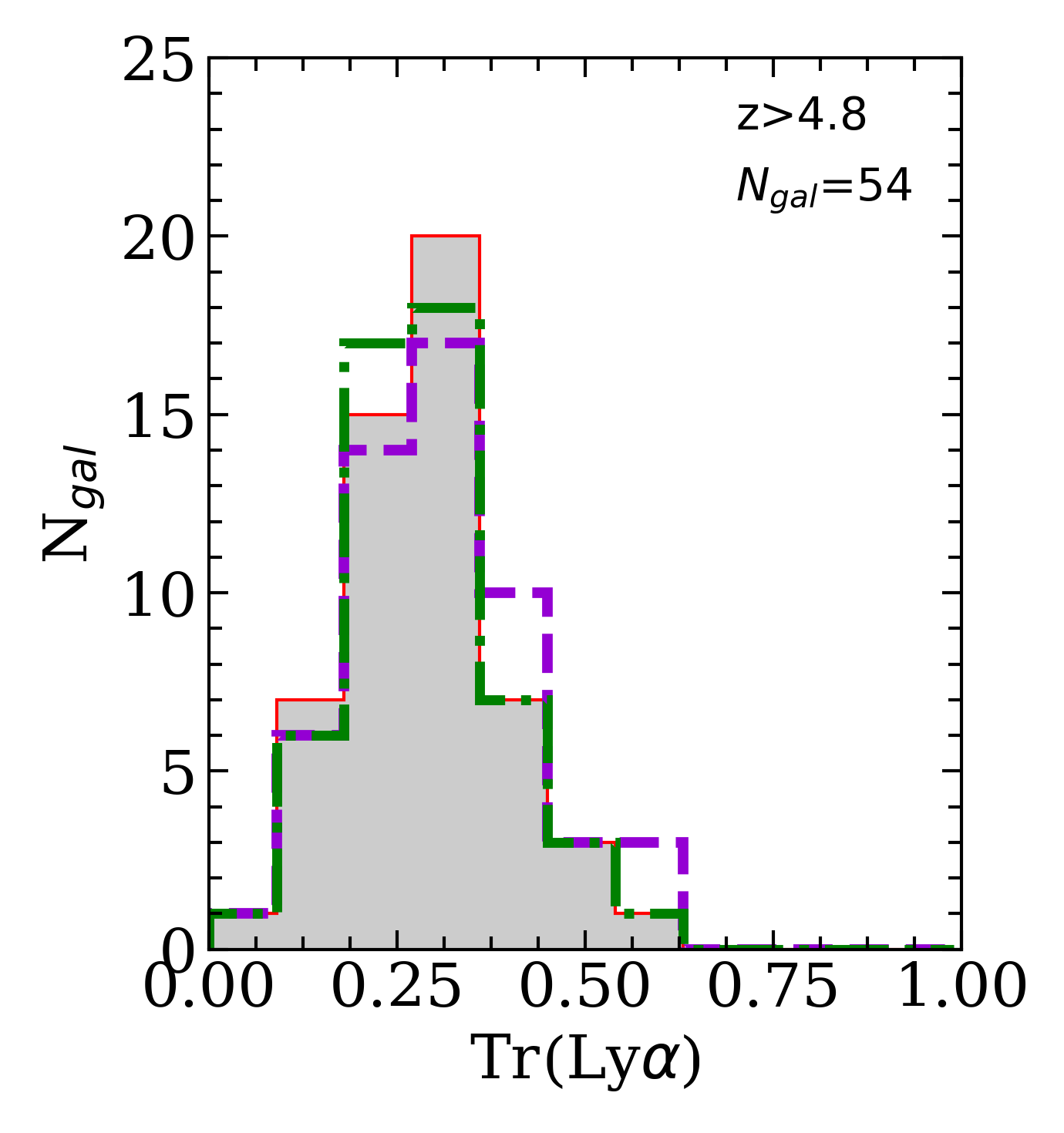}
\includegraphics[width=\hsize]{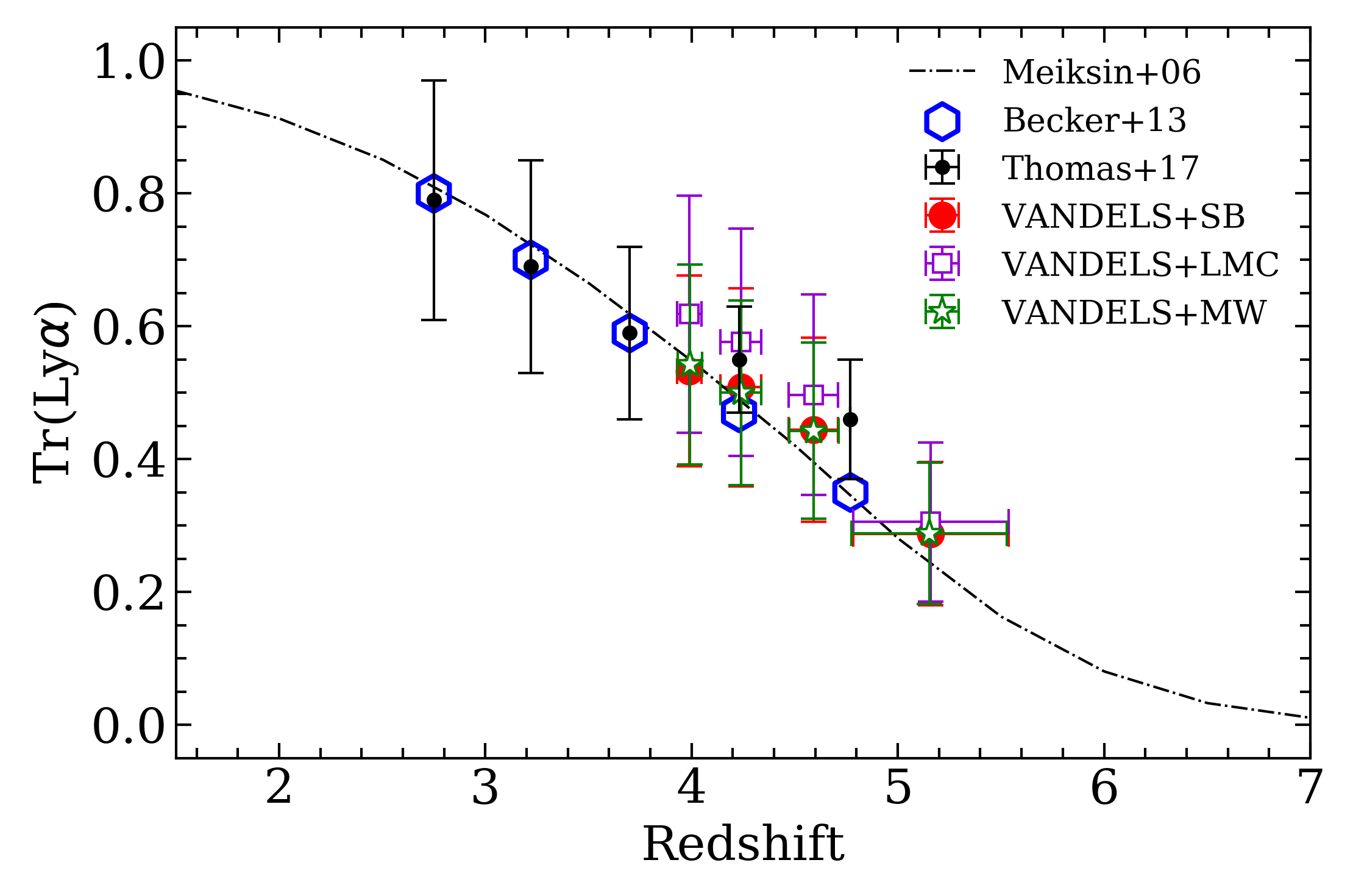}
\caption{Lyman $\alpha$ transmission (Tr(Ly$\alpha$)) as function of redshift. The four top first small plots show the distributions of the transmission in four redshift bins 3.85<z<4.1, 4.1<z<4.4, 4.4<z<4.8 and z>4.8 for each dust prescription. We indicate in each of these plot, the number of galaxies entering in the distribution. The bottom plot display the evolution the transmission with redshift. Our measurements are indicated in red for \textit{SB} and green for \textit{MW} and in violet for \textit{LMC}. We show the measurements from QSO in blue from \cite{Becker13}, from the VIMOS Ultra Deep Survey (VUDS, \cite{Thomas17}, in black) and the theoretical prediction from \cite{Meiksin06} represented with the black dashed line.}
\label{lyatrd}
\end{figure}

Results on the measurement of Tr(Ly$\alpha$) are presented in Fig. \ref{lyatrd} where we show the distributions of the Lyman $\alpha$ transmission in four redshift bins: 3.85<z<4.1; 4.1<z<4.4, 4.4<z<4.8 and at z>4.8. We also display the evolution of this quantity with redshift as compared with previous measurements in the literature (we take uneven binning to ensure a maximum number of galaxies in each bins). Table. \ref{table} provides the measurements for each bin.
We estimated the Ly$\alpha$ transmission in these four redshift bins. Using the SB dust attenuation, this quantity goes from Tr(Ly$\alpha$)=0.55 at z=3.99 with a large standard deviation of 0.14 to Tr(Ly$\alpha$)=0.29 at z=5.15 with a standard deviation of 0.11. The standard error of the mean is small and is below 0.02 at any redshift. The measurements using the MW dust curve are giving similar measurements. This measurements are similar to measurements done with QSOs at similar redshift. \cite{Becker13} measured Tr(Ly$\alpha$)=0.59 at z=3.70 and Tr(Ly$\alpha$)$\sim$0.35 at z$\sim$4.8. This shows that even at high redshift we are able to reproduce equivalent measurements with galaxies. Comparing our results to theoretical prediction, we find that we are in good agreement with the M06 models that predicts Tr(Ly$\alpha$)=0.39 at z=4.6 and Tr(Ly$\alpha$)=0.25 at z=5.15. We note that our measurements are in partial disagreement with our previous measurements at z>4 from the VUDS galaxies. At z=4.23, the difference from our previous measurement is more than 10\% and reaches 20\% at z>4.5. Nevertheless, as reported in \cite{Thomas17}, this high values of Tr(Ly$\alpha$) could actually be corrected limiting the E(B-V) to low values. Thus, the method we employed in the present paper with an estimation of the E(B-V) value before the spectral fit seems to correct for this degeneracy. 

More importantly we report a large standard deviation of Tr(Ly$\alpha$) for all our measured points. It goes from 0.14 at z=3.99 to 0.11 at z=5.15. This is in good agreement with our previous study and confirms that IGM should be treated as free parameter during fitting process. Surprisingly, we find that the measurement using the LMC prescription are above the other ones with a difference that peaks at +0.08 at z=4.23 while the highest redshift point is in good agreement with the other dust solutions. We try to investigate these differences in the next section.

\section{Averaged spectra}
\label{stacks}

Finally, we have a look at averaged spectra of the population of our selected galaxies. We build two stack spectra that are constructed based on the IGM transmission as measured in our VANDELS galaxies: one where we select all the galaxies with a transmission higher than the mean curve given by M06, and one where all the galaxies have a transmission lower than the mean curve. Each stack spectrum is constructed using the \textit{specstack}\footnote{https://specstack.readthedocs.io/en/latest/} program \citep{specstack} that works as follows. For a given stack we de-redshift all the individual spectra and normalise them in a region red-ward of the Ly$\alpha$ line free of emission or absorption lines, in this case between the SiII($\lambda1260$) and OI($\lambda1303$) absorption line. Then we re-grid all the spectra in a common wavelength grid. Finally, at a given wavelength, we compute the mean of all the fluxes using a sigma-clipping method (at 3$\sigma$). 
\begin{figure}[h!]
\centering
\includegraphics[width=\hsize]{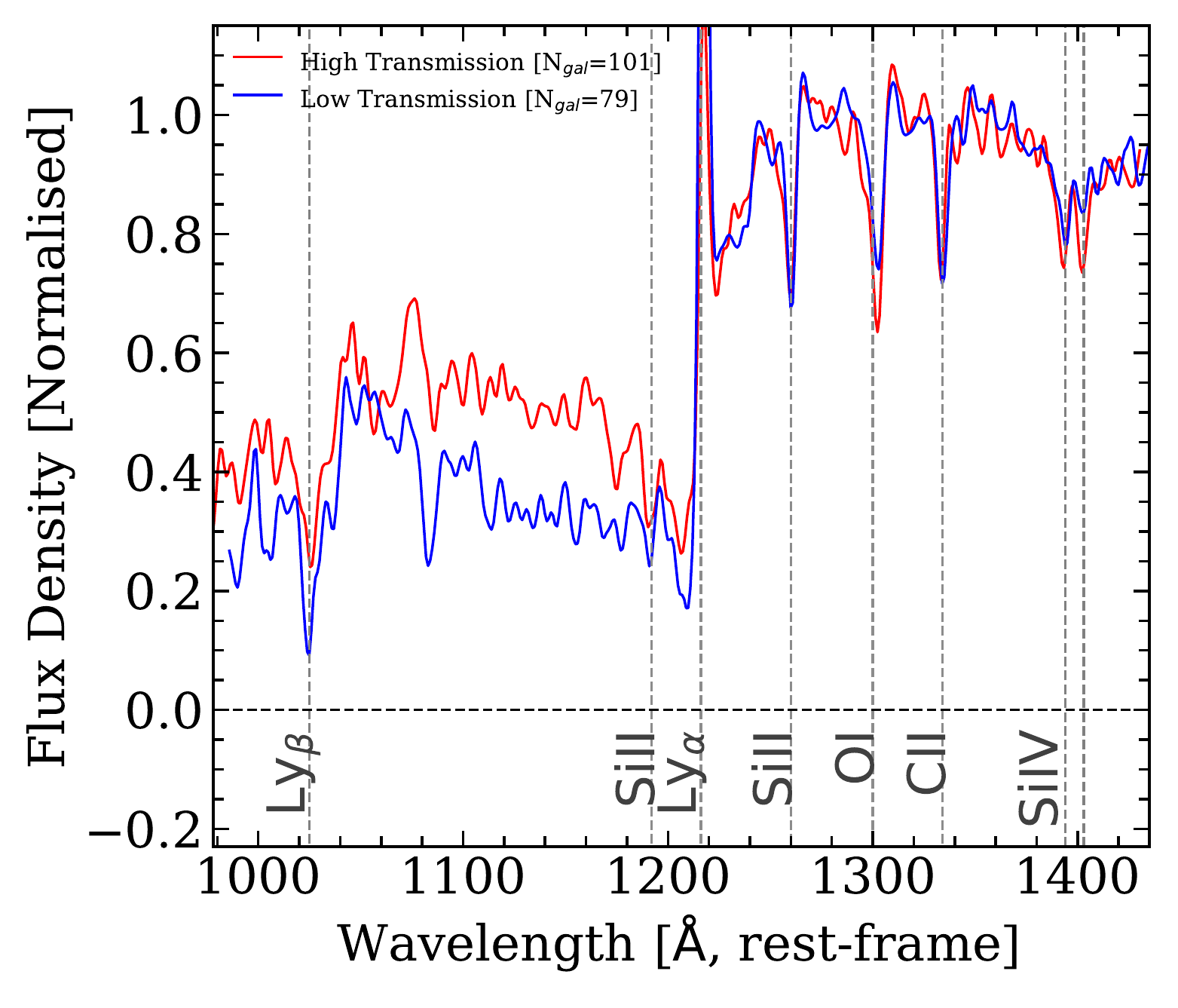}
\caption{Averaged spectra. We display two stack spectra. In blue we show the average of all the spectra (79) with IGM transmission lower than the mean. The mean redshift is these spectrum $\sim$4.36. In red we show the average of all the spectra (101) with IGM transmission higher than the mean, with a mean redshift of $\sim$4.60. The stack spectra have been made with the \textit{specstack} program \citep{specstack}.}
\label{stackfig}
\end{figure}

The averaged spectra are presented in Fig. \ref{stackfig} (using the fits made with SB). This figure shows that below Ly$\alpha$ at 1215$\mathrm{\AA}$, there is a non-negligible variation in flux. The low-transmission stacked spectra are on average $\sim$30\% dimmer than the stacked spectra with high IGM transmission. This means that at z$>$4, the standard deviation of the IGM is very important, and that IGM transmission should be treated as a free parameter when studying galaxy at such high redshift. It is also worth mentioning that the average redshift of the two stacks is slightly different. For the stacks with IGM below than the mean, the average redshift is $z\sim4.36$ while it is $z\sim4.60$ for the galaxies with an IGM higher than the mean. Consequently, the difference might be even higher than what we measure here. Finally, the figure shows that the spectra beyond the Ly$\alpha$ line are very similar. Few absorption lines are stronger in the case of high IGM transmission (e.g. OI and SiIV) but others present similar strength (e.g. SiII and CII) it is therefore delicate to conclude on this aspect.

\section{Discussion}
\label{disc}

\subsection{Flag system and flux calibration}
\begin{figure}[h!]
\centering
\includegraphics[width=\hsize]{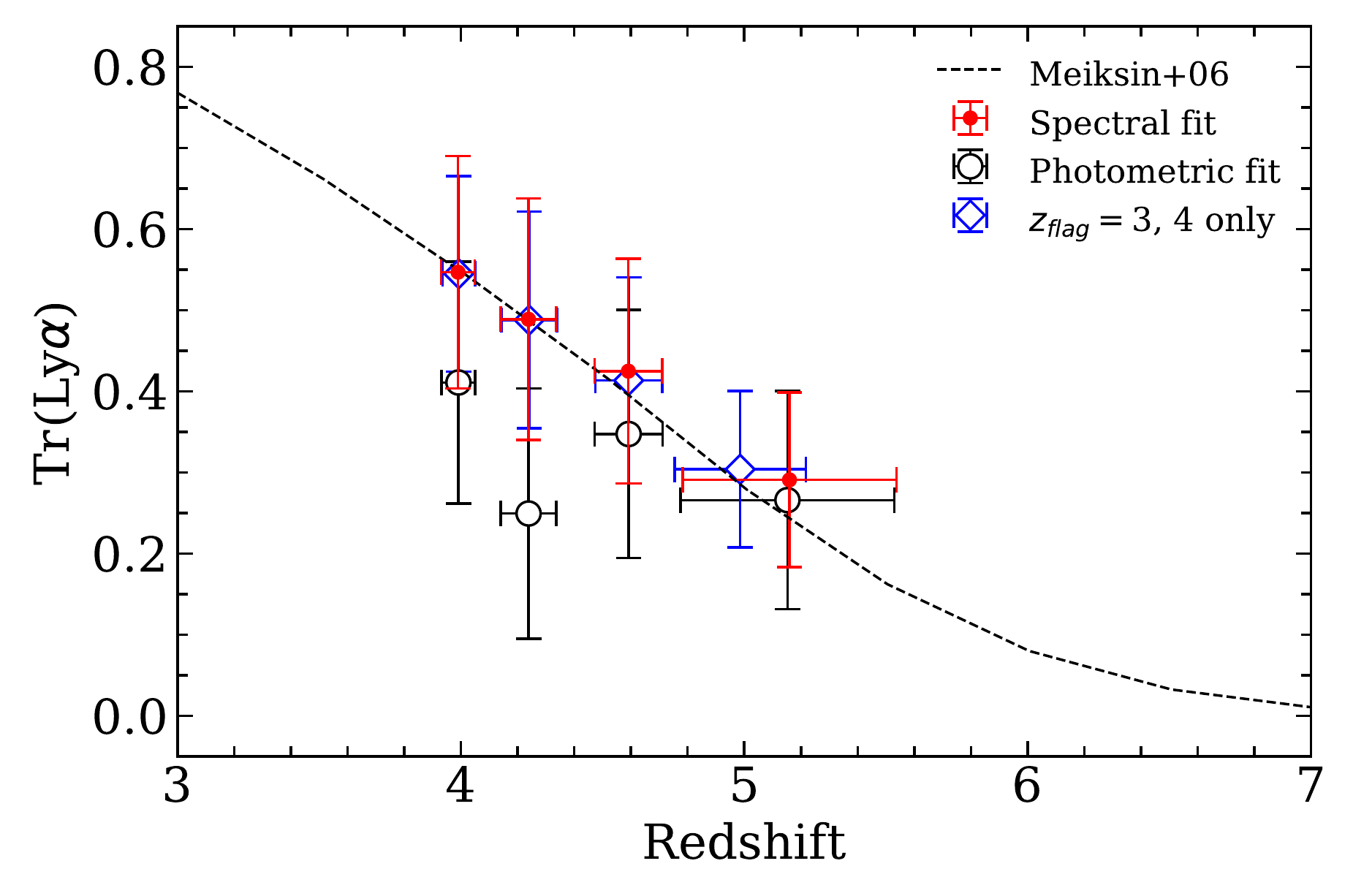}
\caption{Intergalactic medium transmission Tr(Ly$\alpha$) as function of redshift from different estimation. In red we show our final results, as presented in Fig.\ref{lyatrd}, in blue we show the evolution of Tr(Ly$\alpha$) for galaxies with a redshift flag of 3 or 4 only and in black we show the results out of the photometric fit only.}
\label{tests}
\end{figure}

As mentioned in Sect. \ref{Data} we did not take into account the flag system when selecting our galaxies. However we checked if our result are impacted by the presence of lower quality redshift measurements that could potentially indicate presence of low-redshift interlopers. We removed the redshift flags 1 and 9 and results are reported in Table \ref{table} and displayed in Fig.~\ref{tests}. The only notable difference is the last point that is at a slightly lower redshift at z=4.98 instead of z=5.15, indicating that the highest measured redshift have a lower quality than the lower redshift sample. This is because the redshift is often measured due to the presence of the Ly$\alpha$ which leads to a redshift quality flag of 9. For the other points the changes in Tr(Ly$\alpha$) is less than 0.01, which represent less than 2\%  of difference. We conclude that including redshift flag 1, 2 and 9 have almost no impact on the global result on our study.\\

As reported in \citet{Pent18} the very bluest part of the VANDELS spectra was suffering from a systematic mismatch with the broad-band photometry available for the sources. The underlying cause for this is still under investigation but for the moment (and at the time of DR1 and 2)  we have implemented an empirically derived correction to the spectra. This effect could in principle be relevant  for the objects belonging to the first redshift bin. For this reason we repeated the same measurements in the two first bins using uncorrected spectra and measure an IGM transmission of 0.55 at z=3.99 with a standard deviation of 0.16 and 0.51 at z=4.23 with a standard deviation 0.16 as well. This shows that the spectral corrections effects are negligible and lower than 5\%.

\subsection{Discussion on the method}
Another measurement we have done is the measurement of the IGM transmission from the first pass photometric fitting performed in Sect.~\ref{sec_dust} (with free dust extinction parameter). Results are reported in Fig.~\ref{tests} and Tab.~\ref{table}. As expected the measurements are in strong disagreement with the results from the spectral fit. The difference is on average -0.11 towards low Tr(Ly$\alpha$) values. This difference can reach up to 0.24 at z=4.23. This shows that the use of photometric data alone to constrain the IGM transmission is not efficient. Photometric data are less numerous and we have access to less bands to constrain the fit. Indeed, photometry provides us with a data-point every 500 or 1000\AA~. Spectroscopy, to the contrary, brings much more constraints with one data-point every $\sim$4\AA.\\

We finally test the difference in dust extinction estimation (E(B-V), using SB) and Tr(Ly$\alpha$) if we do not fix the dust extinction during the spectral fit. 
Leaving it free, the dust extinction that we measure increases with respect to the photometric fitting of Sect.\ref{sec_dust}. The measurement of E(B-V) give 0.12, 0.13, 0.15 and 0.12 at z=3.99, 4.23, 4.59 and z=5.15. While we are still in the dispersion, this corresponds to a change of 10\% at z=4.05 and more than 50\% at z=5.15. When looking at Tr(Ly$\alpha$), the measurements is slightly different from the main results of our paper. The first point at z=3.99 remains the same while the second and third measurements are higher when leaving the dust free with Tr(Ly$\alpha$)=0.51 at z=4.23 and Tr(Ly$\alpha$)=0.44 at z=4.59. This makes a difference of between 0.03 and 0.02, respectively. The strongest difference is for the last point where the $\Delta$Tr(Ly$\alpha$)=0.05. These behavior is expected. If the dust content goes toward higher E(B-V) values (i.e. more extinction), the IGM transmission must compensate this extinction going toward higher Tr(Ly$\alpha$) values (higher transmission). This behaviour was already noted in our previous study. \\

\subsection{IGM template resampling}
\begin{figure}[h!]
\centering
\includegraphics[width=\hsize]{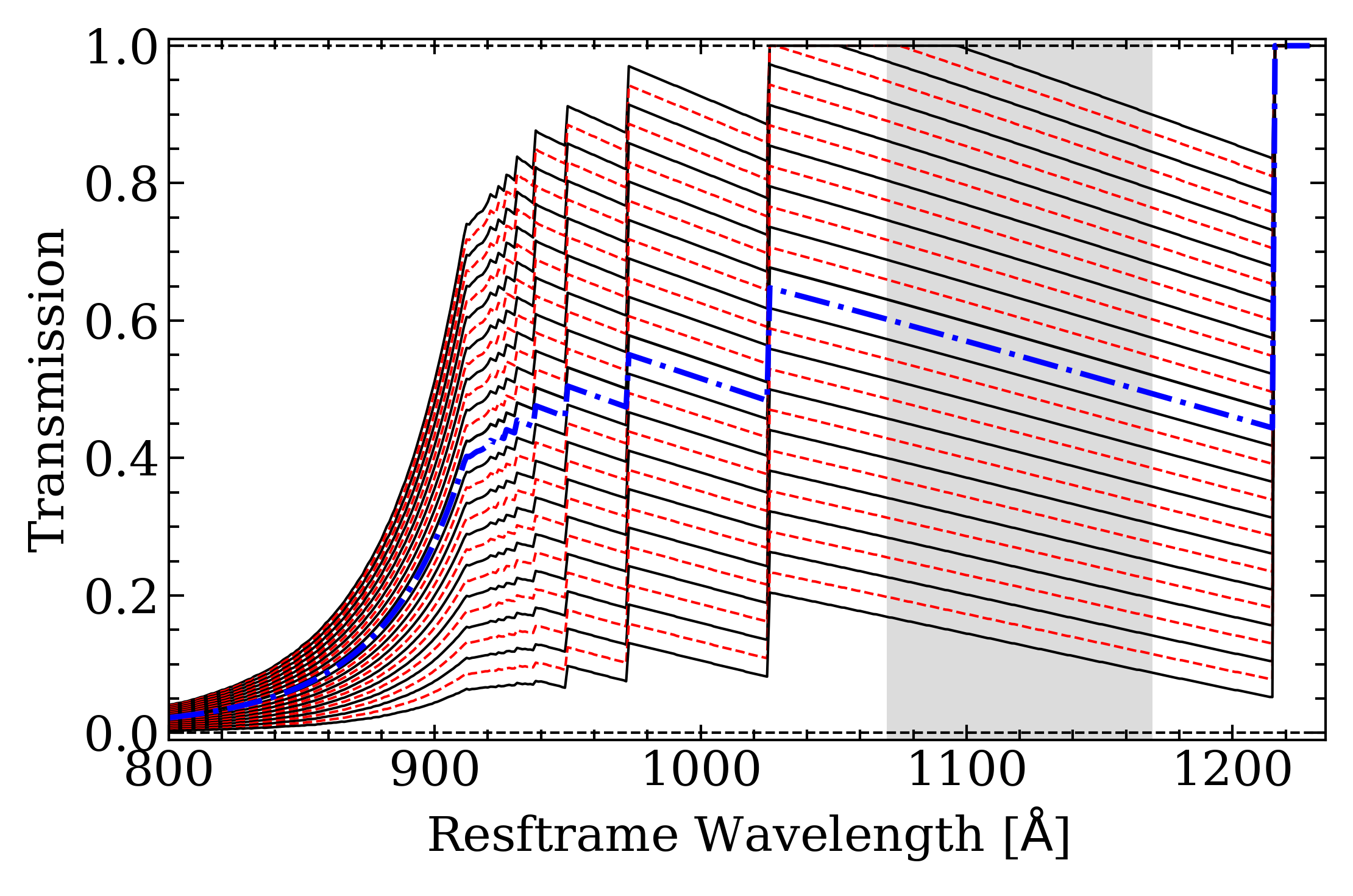}
\includegraphics[width=\hsize]{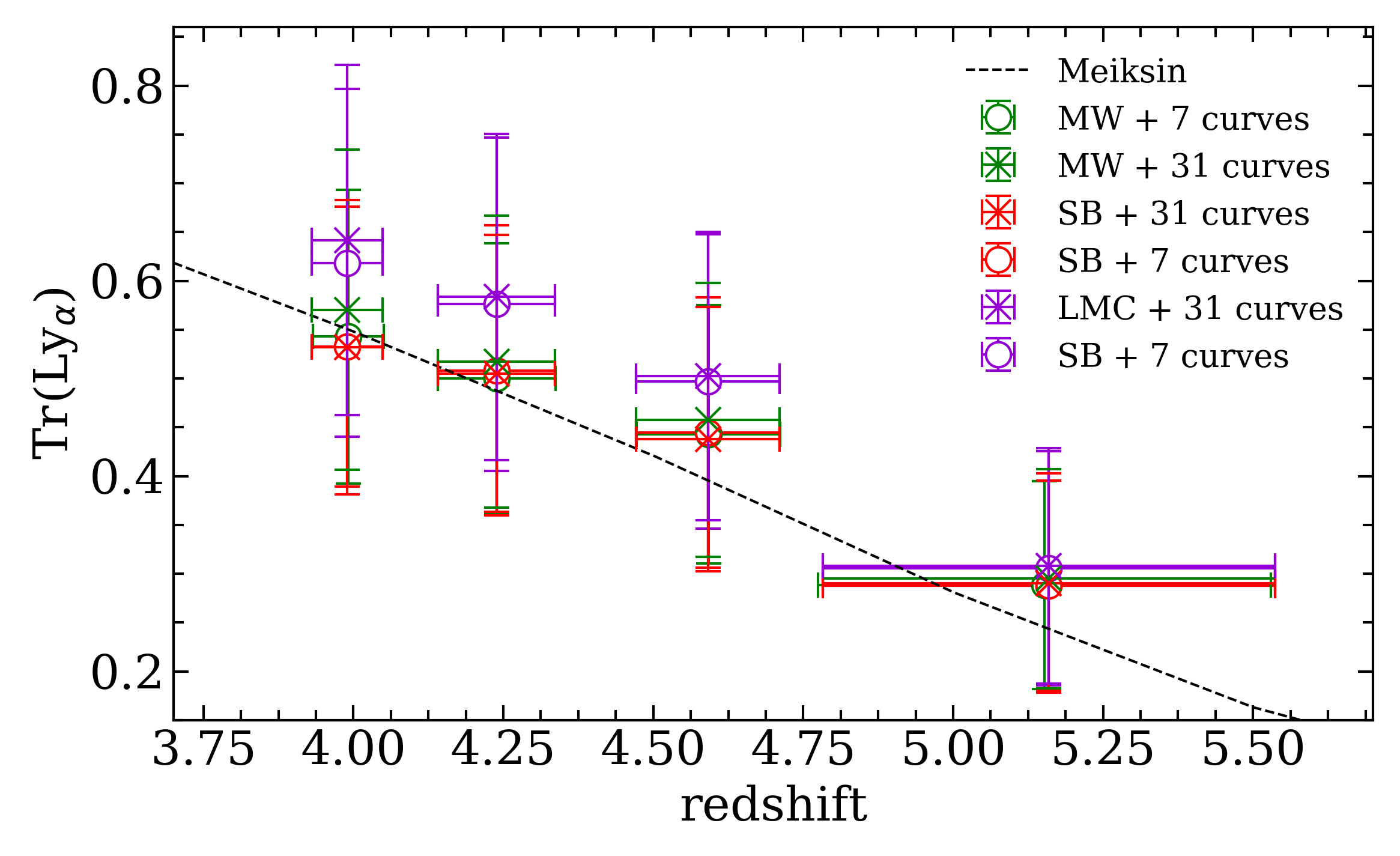}
\caption{\textit{Top:} Intergalactic medium transmission at z=4 in the case where we finely sample the IGM templates and consider 31 transmission curves instead of 7. The blue curve shows the average M06 prescription and the grey region locates the region where we measure Tr(Ly$_\alpha$). \textit{Bottom:} Comparison of the Tr(Ly$\alpha$) in the case where we use 7 curves or 31 curves.}
\label{mega}
\end{figure}

In the work presented previously we used 7 IGM transmission curves at any redshift. We want to know now if the sampling of our prescription has an influence on the final measurements. To this aim, we created a new IGM prescription, not composed of 7 possible transmissions, but of 31 transmission templates. We keep the same range but add intermediate curves, each at multiples of 0.1$\sigma$ from 0.1 to 1.5$\sigma$. The transmission curves at z=4.0 can be seen in Fig.\ref{mega} (top).

Using this fine-sampled prescription we recompute the dust and the IGM transmission using all the three dust prescription, the results are displayed in Fig.~\ref{mega} where we compare the results from the fit with the 7-curves prescription and the ones where we use the 31-curves version of the IGM prescription. This comparison shows that the difference is minimal. Using the SB dust attenuation the use of both prescription has no effect and the difference is less than 0.1\%. For the two other prescription the main difference is for the point at the lowest redshift where the difference reaches 4\% for the LMC and 5\% for the MW prescription. We can conclude that the prescription with 7-curves seems to be detailed enough and adding more curves does not substantially change the results.

\section{Conclusion}
This paper reports the study of the intergalactic medium transmission Tr(Ly$\alpha$) at z>4. We measured the IGM transmission from the spectra of 281 galaxies coming from the VANDELS public survey carried out by the VIMOS instrument at the VLT. Galaxies have been observed up to $\sim$80h, thus providing unprecedented spectral depth. Using a previously published IGM transmission prescription for template fitting studies we used the SPARTAN fitting tool to compute the IGM transmission. We summarize our results below:

\begin{itemize}
\item In order to tackle the dust-IGM degeneracy that was discovered in a previous study, we estimated first the dust content in our galaxies with a pure photometric fitting technique. We estimated the mean E(B-V) at z>4 and found that it ranges from 0.11 at z=3.99 to 0.08 at z=5.15. These measurements are similar to previous measurements reported in the literature.
\item Using individual measurements of E(B-V) as constraint, we use the SPARTAN software to perform the spectral fitting of our galaxies. From this fitting we extract the values of Tr(Ly$\alpha$) at various redshift. It decreases from Tr(Ly$\alpha$)=0.53 at z=3.99 to Tr(Ly$\alpha$)=0.28 at z=5.15. These results match very well the measurements from QSOs study and from theoretical predictions. This reinforces the fact that high redshift galaxies can be used to estimate the IGM.
\item Even more importantly the $1\sigma$ scatter of Tr(Ly$\alpha$) is large at any redshift. It is higher than 0.1 and equivalent to the standard deviation reported from QSOs data.
\item As expected we find that the IGM transmission measurements are sensitive to the choice of dust attenuation prescription. 
\item We test whether our results are sensitive to the redshift flag system in place in VANDELS and find that the differences are minimal. 
\item Due to a lack of observational constraints, the measurements coming from a pure photometric fitting are not able to reproduce the results from the spectral fitting and from the literature.
\item Finally, we compute the IGM transmission leaving the dust extinction as a free parameter and we confirm the presence of dust/IGM degeneracy. In that case, the dust extinction goes toward higher value that are in tension with measure from the literature. This is then compensated by an higher IGM transmission. 
\end{itemize}

Finally, it is worth reminding that they are multiple IGM models in the literature and they should be tested against real data. High-redshift data samples are getting big enough to make statistically significant tests of these models. This will be studied in a paper in preparation.

\begin{acknowledgements}
The authors wish to thank the referee we provided us with very insightful comments that improved a lot the paper.
\end{acknowledgements}

\bibliographystyle{aa}
\bibliography{IGM_vandels.bib}

\begin{appendix}
\section{Reproducibility}
\label{secrep}
The reproducibility aspect had become a crucial aspect of modern research with the use of softwares and codes. Sharing codes and method in paper is as important as sharing results. In this appendix we aim at answering this aspect. Table \ref{Table_repro} lists the availability of the data-related and technique-related aspects of our work. Each point is detailed in the next paragraph.
\begin{table}[h!]
\centering
\caption{Summary of the reproducibility of this work}
\begin{tabular}{cccc}
\hline
 & Public & Partial & Private  \\ 
\hline 
\hline
VANDELS Data  & $\surd$ & $\chi$  & $\chi$  \\ 
SPARTAN-tool  & $\surd$ & $\chi$  & $\chi$ \\
Spectral measurements & $\surd$ & $\chi$  & $\chi$ \\
Results & $\surd$ & $\chi$  & $\chi$ \\
Plotting tool: Photon& $\surd$ & $\chi$  & $\chi$ \\
fits file library: dfitspy & $\surd$ & $\chi$  & $\chi$ \\
\hline 
\end{tabular} 
\label{Table_repro}
\end{table}

\begin{itemize}
\item As presented in Sec. \ref{Data}, the VANDELS survey is a public spectroscopic survey. As such all the data are already publicly available and freely available from the ESO archive facility\footnote{http://archive.eso.org/cms.html}.
\item The SPARTAN-tool is available on GITHUB and comes with all the inputs needed to make the code run. The version released at this moment allows separate fit on the photometry and on the spectroscopy, as used in this paper. The version used in this paper is version 0.4.4\footnote{https://astrom-tom.github.io/SPARTAN/build/html/index.html}. The final version will be presented in a paper in preparation (Thomas et al, in prep). 
\item In Addition, the main python packages used during this work are public: catalog query module \textit{catscii} (v1.2, \citealt{catscii}), catalogue matching algorithm \textit{catmatch} (v1.3 \citealt{catmatch}), our fits display library \textit{dfitspy} (v19.3.4, \citealt{dfitspy}), the spectrum stacking program \textit{specstack} (v19.4, \citealt{specstack}) and our plotting tool, \textit{Photon} (v0.3.2, \citealt{photon}). They are all available in the main python package index repository (pypi).
\end{itemize}
\end{appendix}
\end{document}